\documentclass{aa}

\usepackage{amsmath}
\usepackage{siunitx}
\usepackage{xfrac}
\usepackage{natbib}
\usepackage{xcolor}
\usepackage{graphicx}
\usepackage{grffile}
\usepackage{booktabs}
\usepackage{csquotes}
\usepackage{subcaption}
\usepackage{multirow,array}
\makeatletter
  \renewcommand*\aa@pageof{, page \thepage{} of \pageref*{LastPage}}
\makeatother
\usepackage{hyperref}

\hypersetup{pdfpagemode = {UseNone},
            pdftitle = {Binary Evolution},
            pdfcreator = {\LaTeX},
            pdfproducer = {pdfeTeX-0.\the\pdftexversion\pdftexrevision},
            pdfauthor = {Anna Penzlin, Wilhelm Kley},
            pdfsubject = {},
            pdfview = {FitH},
            pdfstartview = {FitH},
            colorlinks = {true},
            linkcolor = [rgb]{0,0.35,0.7},
            citecolor = [rgb]{0,0.35,0.7},
            filecolor = [rgb]{0.61,0,0},
            urlcolor = [rgb]{0.61,0,0},
           }

\usepackage[normalem]{ulem}

\usepackage{bm}
\renewcommand{\vec}[1]{\bm{\mathrm{#1}}}

\newcommand{\jscrit}{j_\mathrm{s,crit}}
\newcommand{\js}{j_\mathrm{s}}
\newcommand{\Tbin}{T_\mathrm{bin}}

\newcommand{\todo}[1]{\textcolor{red}{\bf (todo: #1)}}

\colorlet{darkgreen}{green!60!black}

\colorlet{darkpink}{magenta!90!black}

\colorlet{darkorange}{orange!80!black}

\newcommand{\beq}{\begin{equation}}
\newcommand{\eeq}{\end{equation}}
\newcommand{\bea}{\begin{eqnarray}}
\newcommand{\eea}{\end{eqnarray}}

\begin{document}

\title{Binary orbital evolution driven by a circumbinary disc}

\author{Anna B.~T.~Penzlin,
        Wilhelm Kley\thanks{deceased},
	Hugo Audiffren \and
	Christoph M.~Schäfer
    }

\institute{
Institut f\"ur Astronomie und Astrophysik, Universität T\"ubingen,
Auf der Morgenstelle 10, D-72076, Germany\\
\email{anna.penzlin@uni-tuebingen.de}\\
}
\date{}

\abstract
{
The question whether the interaction of a circumbinary disc with the central binary system leads to shrinking or expansion of the binary orbit has attracted
considerable interest as it impacts the evolution of binary black holes and stellar binary stars in their formation phase.
We performed two-dimensional hydrodynamical simulations of circumbinary discs
for a large parameter set of disc viscosities and thicknesses and two different binary
mass ratios for binaries on circular orbits. For those we measured carefully the net angular momentum and mass transfer between disc and binary system, and evaluate the normalised specific angular momentum accretion, $\js$.
This is compared to the theoretical, critical specific angular momentum change $\jscrit$ that separates contracting from expanding cases which depends on the the binary's mass ratio and the relative accretion onto the two stars.
Using finite and infinite disc models we show that the inferred binary evolution is very similar for both setups and confirm
that $j_\mathrm{s}$ can be measured accurately with cylindrical simulations that do not include the central binary.
However, to obtain the relative accretion onto the stars for non-equal mass binaries, simulations that cover the whole domain including the binary are required.
We find that for thick discs with aspect ratio $h=0.1$ the binaries expand for all viscosities, while discs with $h=0.05$ lead to an expansion only for
larger viscosities with $\alpha$ exceeding $\sim 0.005$.
Overall, the regime of binary expansion extends to a much wider parameter space than previously anticipated,
but for thin, low viscosity discs the orbits shrink.
}


\authorrunning{Penzlin et.~al}

\maketitle

\section{Introduction}\label{sec:intro}
Circumbinary (CB) discs orbiting two gravitating objects exist over a wide range of spatial scales, starting from spectroscopic binary stars, then to large binary systems like GG Tau, and finally all the way to galactic discs around binary black holes.
The binary star and the disc resemble a coupled system that exchanges mass and angular momentum between its components leading possibly to a
secular evolution of the binary's orbit, i.e.\ changes in its semi-major axis and eccentricity. Due its gravitational action the binary
transfers angular momentum to the disc. As a consequence the disc recedes from the binary opening a central cavity, while the binary reacts  with a shrinkage of its orbit accompanied by an increase in its eccentricity \citep{1991ApJ...370L..35A}.
In the context of black holes such a disc induced orbital evolution
was already suggested by \citet{1980Natur.287..307B} to overcome the so called 'final parsec problem' \citep{2003AIPC..686..201M},
i.e. the necessary contraction of the orbit such that gravitational wave emission can set in.
In the context of binary star formation a CB-disc will be formed that determines the binary orbital evolution similarly.

Given the binary's gravitational torque acting on the CB-disc and its own angular momentum loss,
it was always assumed that the binary can only react with a shrinkage of the orbit. This
point of view was questioned by \citet{2017Miranda} who argued that for certain binary parameter
the mass accretion and, hence, accreted angular momentum onto the central binary 
can counterbalance and even overcome the disc torques
leading to an expansion of the orbit. This has led to an increased activity to study the still controversial issue of binary orbit evolution in more detail using
various numerical approaches.

In their simulations \citet{2017Miranda} used a grid-code operating in cylindrical coordinates, where the binary stars orbited
inside of the computational domain, and their evolution was calculated by monitoring mass and angular momentum balance of the surrounding
disc. Concerns about the impact of the grid's inner radius on the outcome were resolved by simulations that covered the whole domain
including the stars. Using a moving mesh code, \citet{2019Munoz} presented simulations that covered the whole domain and clearly
demonstrated that the orbits of binary stars can indeed expand due to the action of a CB-disc, a result which was confirmed by
\citet{2019ApJ...875...66M} using a Cartesian grid.
The topic of binary evolution was treated subsequently by different groups \citep[e.g.][]{2020ApJ...889..114M, 2020ApJ...901...25D, 2020ApJ...900...43T, 2020A&A...641A..64H, 2021ApJ...909L..13Z} using different numerical approaches.

The first works \citep{2017Miranda,2019Munoz,2019ApJ...875...66M} 
focused primarily on equal mass binaries surrounded by discs with high viscosity ($\alpha=0.1$) and larger thickness ($H/r=0.1$) because those reach a quasi-stationary state earlier, and are easier to simulate.
Taking into account the disc's backreaction on the binary \citet{2019Munoz} shows that initially circular binaries remain circular during the evolution.
Later, \citet{2020ApJ...889..114M} performed a larger parameter survey where they varied the viscous $\alpha$ parameter from 0.01 to 0.1 and the binary star mass ratio $M_2/M_1$ from 0.1 to 1.0.
They found that binaries with larger mass ratio ($M_2/M_1>0.3$) expand independently of the used $\alpha$-value.

The quoted studies agree that for an equal mass binary system inside a thick disc with $H/r \sim 0.1$ and a large viscosity,
the added angular momentum of the disc is sufficient to allow for an expanding binary orbit.
However, \cite{2020ApJ...900...43T} applying a constant kinematic viscosity
, varied the disc thickness and found that
for thinner discs with $H/r \lesssim 5\%$ the trend reverses and the binary orbit shrinks again.
According to \cite{2020ApJ...900...43T} the reason for the change of this behaviour lies in the fact that in thinner discs the cavity is deeper and 
more material remains near the cavity walls exerting a negative torque on the binary.

In general, disc parameters such as scale height and viscosity influence shape and size of the eccentric inner disc cavity
\citep{2017Thun,2018Thun,2020Hirsh,2021A&A...645A..68P}, which in turn impacts the transfer of angular momentum between disc and binary.
It is still an open question how the binary's orbit will evolve for ranges of physical disc parameter towards low viscosity and disc heights.
In this paper we present our results of a wider parameter study. 
In particular, we are interested in thin discs resembling a protoplanetary environment.

Using two binary mass ratios we compute the evolution of
a CB-disc varying the viscosity and aspect ratio independently and calculate the binary's orbital evolution.
We will present models using an $\alpha$-type disc viscosity, as well as constant viscosity models.
Our study is restricted to circular binaries to reduced the changing orbit parameters to the semi-major axis and make long model runs for low viscosities feasible.

We confirm the statement made by \cite{2017Miranda} 
that for circular binaries their change in semi-major axis can be accurately obtained from simulations using a cylindrical grid
that surrounds the binary, if all the relevant fluxes across the boundaries are measured,
and for a known mass accretion ratio between the stars.
Using infinite as well as finite disc profiles we show that the inferred binary evolution is very similar for both setups \citep[see also][]{2020ApJ...889..114M}.
We expand the parameter range to thinner as well as less viscous discs, relevant for protoplanetary scenarios, and determine the transition between
expanding and shrinking binary orbits.
To validate our results, we use two grid codes utilising different numerical algorithms,
\texttt{RH2D} \citep{1989A&A...208...98K,1999MNRAS.303..696K} and \texttt{PLUTO} \citep{2007Mignone, 2017Thun}.
To estimate directly the mass accretion onto the individual stars, we used \texttt{RH2D} on a Cartesian grid and additionally Smoothed Particle Hydrodynamics (SPH) simulations \citep{2016A&A...590A..19S} (see Appendix \ref{sec:appC_facc}).

In Section \ref{sec:theory} we discuss the angular momentum evolution of disc and binary from a theoretical standpoint,
and define a critical specific angular momentum accretion rate that separates inspirals from expansions.
We present our physical and numerical setup for the two simulation codes in Section \ref{sec:model} and 
display in Section \ref{sec:standard} the results of the simulations for one standard model.
Section \ref{sec:parameters} gives an overview of the results for our parameter sets in viscosity, scale height and binary mass ratio.
Finally, we discuss and summarise our findings in Sections \ref{sec:discussion}.

\section{Angular momentum evolution of disc and binary}
\label{sec:theory}
Before presenting our numerical results we briefly discuss the orbital evolution of
the central binary from a theoretical standpoint.
Similar considerations were presented for example by \citet{2019Munoz} and \citet{2021ApJ...921...71D}.
For a binary with total mass $M_\mathrm{b}$, semi-major axis $a$ and eccentricity $e$
the orbital angular momentum is given by
\beq
      J_\mathrm{b} = \mu_\mathrm{b}  \left[ G  M_\mathrm{b} a ( 1 - e^2) \right]^{1/2} \,,
\label{eqn:Jbin}
\eeq
where
\beq
     \mu_\mathrm{b}  = \frac{M_1 M_2}{M_1 + M_2}  = \frac{q}{(1+q)^2}M_\mathrm{b}
 \label{eqn:J_bin}
\eeq
is the reduced mass and $q=M_2/M_1$ is the mass ratio of the binary star with $M_2 \leq M_1$.
The time evolution of the angular momentum involves changes of the four quantities
$M_\mathrm{b}(t), q(t), a(t)$ and $e(t)$, which requires in turn knowledge about four parameters:
the change in total mass, orbital angular momentum, momentum, and a knowledge about how the accreted mass and momentum
is distributed amongst the two stars. 
These distributions can be calculated in simulations that include the binaries and flows onto the binaries in the domain with high resolution, as done for example in \citet{2020ApJ...889..114M} for thick discs. For long runs with low viscosity and thin discs a domain including the binary is currently not feasible.

In this work we will from now on assume $e=0$.
By restricting our study to circular binaries with maximum angular momentum for given masses, any addition
of angular momentum will not raise the eccentricity, but rather lead to an expansion of the orbit that remains circular. 
This is supported by simulations of \citet{2019Munoz} who find only very small increase of binary eccentricity in their simulations. 
Hence, we do not consider any change of the binary eccentricity and set $\dot{e}=0$.
Taking the time derivative of eq.~(\ref{eqn:Jbin}) for circular binaries yields
\beq
      \frac{\dot{J}_\mathrm{b}}{J_\mathrm{b}} = \frac{\dot{q}}{q}\frac{1-q}{1 + q}
     + \frac{3}{2} \frac{\dot{M}_\mathrm{b}}{M_\mathrm{b}}  + \frac{1}{2}  \frac{\dot{a}}{a} \,,
\label{eqn:dot-Jbin}
\eeq
where the change in mass ratio, $\dot{q}/q$ in terms of the mass changes of the individual stars is given by
\beq
\frac{\dot{q}}{q} = \frac{\dot{M}_2}{M_2} - \frac{\dot{M}_1}{M_1} \,.
\label{eqn:q_dot}
\eeq

The change in binary mass is directly given by the mass accretion rate from the disc, i.e. the material that is advected from
the disc onto the stars, $\dot{M}_\mathrm{b} = \dot{M}_\mathrm{adv}$, but for the
change in angular momentum there are two contributions, first the accreted angular momentum from disc, $\dot{J}_\mathrm{adv}$,
which accompanies the mass flux, and secondly the gravitational torque exerted by the disc on the binary, $\dot{J}_\mathrm{grav}$.
In principle, there could also be a viscous angular momentum flux across the inner boundary (positive or negative)
but this turned out to be very small in our simulations and is neglected here.

The angular momentum change of the binary can either go into the orbit and/or the spins of the individual stars.
Here, we assume that the latter contribution is overall small, as shown in \citet{2019Munoz}, and consider only the change of the binary's orbit.
 
\begin{figure}[htb]
    \centering
    \includegraphics[width=0.45\textwidth]{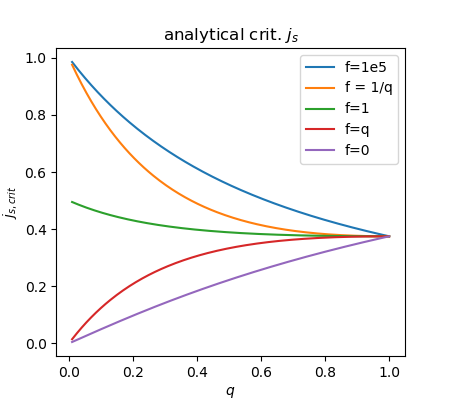}
    \caption{Critical angular momentum transfer to the binary as a function of the binary mass ratio
    for different values of the accretion factor, see eq.~(\ref{eq:js-crit}). If the binary receives angular momentum at a rate larger than $j_\mathrm{s,crit}$ then its orbit expands.
    Displayed are 5 different values for the factor $f$ that enclose all cases between the two limiting cases of $f=0$
    (accretion only on the primary) and $f\rightarrow \infty$ (accretion only on the secondary).}
    \label{fig:ana_js}
\end{figure}

In our typical longterm simulations the stars are not within the computational domain, hence
in order to account for the a priori unknown distribution of mass amongst the two stars,
we introduce the factor $f$.
It denotes the mass distribution between the individual stars and is defined by
\beq
       \dot{M}_2 = f \dot{M}_1 \,,
\eeq
i.e. the secondary receives $f$ times as much mass as the primary.
Then we find with eq.~(\ref{eqn:q_dot})
\beq
       \frac{\dot{q}}{q} = \frac{1+q}{1+f} \, \left(\frac{f}{q}-1\right) \, \frac{\dot{M}_\mathrm{adv}}{M_\mathrm{b}} \,.
\eeq
To rewrite eq.\,(\ref{eqn:dot-Jbin}) in terms of the change in semi-major axis as
we use $\dot{J}_\mathrm{b} = \dot{J}_\mathrm{adv} + \dot{J}_\mathrm{grav}$,
and define the (normalised) {\it specific angular momentum transfer}, $j_\mathrm{s}$, to the binary as
\beq
    \frac{\dot{J}_\mathrm{b}}{\dot{M}_\mathrm{adv}} \frac{M_\mathrm{b}}{J_\mathrm{b}}
      = \frac{\dot{J}_\mathrm{b}}{\dot{M}_\mathrm{adv}}\frac{1}{\sqrt{G M_\mathrm{b}a}} \frac{(q+1)^2}{q} \equiv  j_\mathrm{s} \frac{(q+1)^2}{q} \,. 
\label{eq:js-normalising}
\eeq
The change in semi-major axis of the binary can then be written as
      \beq
    \frac{\dot{a}}{a}  =  \frac{\dot{M}_\mathrm{adv}}{M_\mathrm{b}} \left[ 2\, j_\mathrm{s}  \frac{(q+1)^2}{q} - 3
       - 2 \frac{1-q}{1+f} \, \left(\frac{f}{q}-1\right) \right].
  \label{eqn:adot_bin3}
       \eeq
This equation is equivalent to eq.~(9) in \cite{2020ApJ...889..114M}, where their angular momentum accretion eigenvalue is given by $l_0 = j_s a^2 \Omega_\mathrm{b}$
and their 'preferential accretion ratio' is $\eta \equiv \dot{M_2}/M_\mathrm{b} = f/(1+f)$.
A similar expression was given by \citet{2021ApJ...921...71D}.
To find the critical $j_\mathrm{s}$, at which the orbit changes from contracting to expanding, we set $\dot{a} = 0$
and obtain
\beq
  j_\mathrm{s,crit} = \frac{2+q}{2(1+q)^2} - \frac{1 - q}{(1 + f)(1+q)} \,.
 \label{eq:js-crit}
\eeq
If the specific angular momentum transfer is larger than $\jscrit$ the binary orbit will expand, while it will contract for $j_\mathrm{s}$ smaller
than the critical value.
For equal mass binaries with $q=1$, only the first term remains and equals 3/8, identical to the value found by \cite{2017Miranda}, and used later by \citet{2020ApJ...900...43T}.
In Fig.~\ref{fig:ana_js} the critical values of the angular momentum transfer, $\jscrit$, are shown as a function of the mass ratio, $q$,
for selected sample values for the mass accretion ratio, $f$, between the stars. All values of $\jscrit$ are between 0 and 1, for any value of $q$ or $f$. As discussed above, for $q \rightarrow 1$ all curves converge to $\jscrit = 3/8$.
In the case that both stars accrete the same mass, $f=1$, the critical $\jscrit$ lies always above $3/8$.
It becomes  maximal if the secondary star would accrete all mass, $f \rightarrow \infty$.

In the following we present the results of our hydrodynamical simulations to analyse for which binary and disc parameter
the specific angular momentum accretion is larger than $\jscrit$ and thereby sufficient to lead to an expanding binary orbit.

\section{Modelling}\label{sec:model}
In this section we describe first our physical setup, then the numerical methods and boundary conditions and
finally the measuring procedure of the mass and angular momentum balance.
\subsection{Physical setup}
In our study we restrict ourselves to infinitesimally thin two-dimensional (2D) discs that orbit a central binary.
The plane of the disc and the binary are coplanar.
The system is modelled using polar coordinates with the origin at the centre of mass (COM) of the binary star.
For the simulations we keep the binary's orbit fixed and the disc is not self-gravitating.
To allow for a scale free setup, we adopt a locally isothermal equation of state, with respect to the binary's COM.
The disc is not flared but has a constant disc aspect ratio $h=H/r$, where $H$ is the pressure scale height of the disc at distance $r$.
For the viscosity we assume the classic alpha-ansatz and write for the kinematic viscosity coefficient $\nu = \alpha c_\mathrm{s} H$,
where $c_\mathrm{s}$ is the isothermal sound speed of the gas.

For the initial setup of our problem we follow \citet{2020ApJ...889..114M} and use a finite extent of the initial density
distribution, which has the shape of a circular gas torus around the binary. 
For this purpose we define the two radii $r_\mathrm{in}$ and $r_\mathrm{out}$, as well as the two functions
\beq
 \label{eqn:g_1}
  g_\mathrm{in}(r) = \frac{1}{1 + \exp{[- (r - r_\mathrm{in})/(0.1 \, r_\mathrm{in})]}}
\eeq
and
\beq
 \label{eqn:g_2}
  g_\mathrm{out}(r) = \frac{1}{1 + \exp{[r - r_\mathrm{out}]}} \,.
\eeq
The initial density is then given by
\beq
 \label{eqn:sigma_0}
    \Sigma_0(r)   =  \Sigma_\mathrm{base}(r) \times \left[ 1 - \frac{0.7}{\sqrt r} \right] \times g_\mathrm{in}(r) \times g_\mathrm{out}(r) \,.
\eeq
Here, $\Sigma_\mathrm{base}(r)$ denotes a basic power law profile for the density for which we use
$\Sigma_\mathrm{base}(r) \propto r^{-1/2}$ throughout. The constants are chosen such that
shapes similar to \citet{2020ApJ...889..114M} can be constructed. 
These cut-off functions create a smooth transition to a low density beyond $r_\mathrm{in}$ and $r_\mathrm{out}$. Varying these can be useful start from conditions closer to the convergent state.
An example of such an initial $\Sigma$-profile is displayed below in Fig.~\ref{fig:sigma_evol_standard} for our standard model.

\begin{table}[t]
\centering
\begin{tabular}{|c|c|c|c|c|}
\hline
 $q$  &  $\alpha$  &  $h$ &  $r_\mathrm{in}$  & $r_\mathrm{out}$   \\
\hline
   0.5              &    0.1     &   0.1  &  2.5   &  6.0    \\
\hline
\end{tabular}
\caption{Summary of the physical parameters for the standard model. The values are:
the binary star mass ratio $q = \mathrm{M_2}/\mathrm{M_1}$,
the disc viscosity $\alpha$, aspect ratio $h$, and the disc's initial radial extent $r_\mathrm{in}, r_\mathrm{out}$ (see eqs.~\ref{eqn:g_1} and \ref{eqn:g_2}).
 }
\label{tab:standard}
\end{table}

To have a well defined test case we define a {\it standard model} which has the following parameter:
For the binary we chose $q = {M_2}/ {M_1} = 0.5$, and for the disc $\alpha = 0.1$ and $h = H/r = 0.1$ \citep[like][]{2019Munoz}.
Here, the large viscosity and disc thickness allow for a fast evolution of the disc, and disc matter will
be accreted rapidly onto the binary and spread outwards.

The parameter of the standard model and for the initial surface density are
summarised in Table~\ref{tab:standard}.
In a detailed parameter study we will study the effect of varying the viscosity $\alpha$, and aspect ratio $h$
for two values of the binary mass ratio $q$.

\subsection{Numerics}
For the main part of this paper we performed numerical simulations over very long time scales in order to cover discs with alpha-viscosities down to $10^{-3}$. To allow for a reasonable computational time the binary stars are not included in the domain.
In the Appendix \ref{subsec:f} we present an example of a completely covered inner domain for the standard model to the derive the accretion ratio $f$.

We use two grid-based codes operating on a cylindrical grid stretching from $r_\mathrm{min}=a$ to $r_\mathrm{max}=40{a}$
with 684 logarithmically spaced radial grid cells and 584 azimuthal cells. 
Following our previous study \citep{2018Thun}
the value $r_\mathrm{min}=a$ captures the essential dynamics for discs with a significant inner cavity, and we used this value for our
parameter study.
We will show in section \ref{subsec:rmin} how the choice of $r_\mathrm{min}$ possibly affects our measured accretion of mass, angular momentum and $j_\mathrm{s}$.
Concerning the outer radius, it has to be chosen large enough such that
it does not impact the disc's evolution for large and eccentric cavities. Here, we found $r_\mathrm{max}=40 a$ sufficient.
The simulations are carried out in dimensionless units where the unit of length is the binary separation $a$, the gravitational constant $G=1$,
the binary mass $M=1$ and the binary orbital angular velocity $\Omega_\mathrm{b}=1$.

An important ingredient of grid-based simulations is the necessity of a density floor to prevent negative densities and pressures.
For the standard case with its high aspect ratio and large viscosity the inner region close to the stars will never be very empty such that
a floor density of $\approx 10^{-6}$ of the initial maximum density is sufficient. However, for the parameter studies with very small
$h$ and $\alpha$ the cavities around the binary become very deep and much lower floor values $\approx 10^{-10} - 10^{-8}$ are required, in order
to follow the very small mass and angular momentum accretion onto the binary.

In order to validate our results we first calculated the standard model using two grid codes with different numerics,
\texttt{RH2D} \citep{1989A&A...208...98K} and \texttt{PLUTO} \citep{2007Mignone}. We describe the details of the codes and comparison runs in the appendix \ref{sec:A_code}.

\subsection{Boundary conditions}
The cylindrical grid simulations conserve global angular momentum, at least within the active domain.
In our models the inner boundary (at $r_\mathrm{min}$) is open to angular momentum and mass flow towards the star.
We use an inner radial boundary condition of $\partial\Sigma/\partial r = 0$ for the density,
$\partial \Omega/\partial r = 0$ for the azimuthal velocity, and a 
diode-type radial infall with $v_{r,\mathrm{ghost}} = \min ( v_{r,\mathrm{min}},0)$.
The zero gradient condition for the angular velocity implies a vanishing torque at the inner boundary,  at least for circular flows,
and ensures that the disc is not artificially torqued from inside.
We tested other conditions for $\Omega(r_\mathrm{min})$, such as zero gradients of the angular momentum, but found no differences in the disc's
evolution.
At the outer boundary we use reflecting conditions where the radial velocity is set to zero at the outer boundary, $v_{r,\mathrm{max}}=0$.
For the azimuthal velocity we assume the circular Keplerian velocity around the binary's COM. While the outer boundary is closed to mass flow
angular momentum is nevertheless transmitted through $r_\mathrm{max}$ through viscous torques which is monitored during the evolution.
Additional models resembling an 'infinite' disc setup are presented in Subsection\,\ref{subsec:infinite}.

\subsection{Measuring the angular momentum and mass balance}
To calculated the orbital evolution of the binary, we need to monitor the mass and angular momentum accretion from the disc,
Even though these are losses with respect to the disc, we consider here the gain with respect to the binary star and take them as positive
quantities, denoting them as $\dot{M}_\mathrm{adv}$ and $\dot{J}_\mathrm{adv}$. The subscript 'adv' denotes advection (transport) of the
quantities across the inner radius of the disc. For the mass flux this transport is solely advective and given by
\beq
\dot{M}_\mathrm{adv} =\left. \oint \Sigma r \, |v_r| \, d\phi \right|_{r=r_\mathrm{min}} \,,
\label{eqn:mass_adv}
\eeq
where $v_r$ is the radial velocity of the gas, which can only be negative (or zero) for the used boundary condition, i.e. $v_r ( r_\mathrm{min} ) \leq 0$.
For the angular momentum accretion the advective transport across the inner boundary reads
\beq
\dot{J}_\mathrm{adv} =\left. \oint \Sigma r \, |v_r| \, r v_\phi \,  d\phi \right|_{r=r_\mathrm{min}} \,.
\label{eqn:J_adv}
\eeq
We evaluated the angular momentum and mass flow rate through the inner boundary, typically at $r_\mathrm{min} = a$.
As a consistency test we also measured the flow at different inner radii in the same model ($r = [1.25, 1.5, 2]\,a$).
For the equilibrium state we find the same values, as expected. However, for even smaller inner evaluation radii the circum-secondary
disc can interfere with the calculated advection, depending on the mass ratio. More details on smaller $r_\mathrm{min}$ models are shown below in Sect.~\ref{subsec:rmin}.
The second change of the binary angular momentum is due the disc's gravitational torque
\beq
      \dot{J}_\mathrm{grav} = \sum_{k=1}^{2} \, G M_k \, \int_\mathrm{disc} \Sigma(r, \phi)
      \frac{  \vec{r}_k \times (\vec{r} - \vec{r}_k)}{|\vec{r} - \vec{r}_k|^3}
      r d\phi dr \,.
           \label{eqn:grav}
\eeq
While the advective transport is always positive this last contribution can take both signs.

While the quantities $\dot{M}_\mathrm{adv}$ and $\dot{J}_\mathrm{adv} + \dot{J}_\mathrm{grav}$ are sufficient to evaluate the orbital evolution for
circular binaries, it is very useful to have a simple alternative measuring instrument. 
This is accomplished by simply monitoring the total disc mass, $M_\mathrm{disc}$, and angular momentum $J_\mathrm{disc}$ within the computational domain.
As the outer boundary at $r_\mathrm{max}$ is closed $- \dot{M}_\mathrm{disc} =  \dot{M}_\mathrm{adv}$.
For the angular momentum there is an additional viscous loss channel through the outer boundary 
given by
\beq
     \dot{J}_\mathrm{visc} = \left. \int_0^{2\pi} \left( \Sigma \nu r^2 T_{r \phi} d \phi \right) \right|_{r=r_\mathrm{max}}
\eeq
with the viscous stress tensor
\beq
    T_{r \phi} = r \frac{\partial \Omega}{\partial r} + \frac{1}{r} \frac{\partial v_r}{\partial \phi} \,.
\eeq
Written as such, the sign of $\dot{J}_\mathrm{visc}$ denotes either gains ($>0$) or losses from the disc ($<0$).
Thus the angular momentum can be compared using: $-\dot{J}_\mathrm{disc}$=$\dot{J}_\mathrm{adv} + \dot{J}_\mathrm{grav}$ + $\dot{J}_\mathrm{visc}$.
This monitoring of the global disc properties throughout the simulations allows for a very useful
consistency check on overall mass and angular momentum conservation of the numerics.

\begin{figure}[htb]
    \centering
    \includegraphics[width=0.45\textwidth]{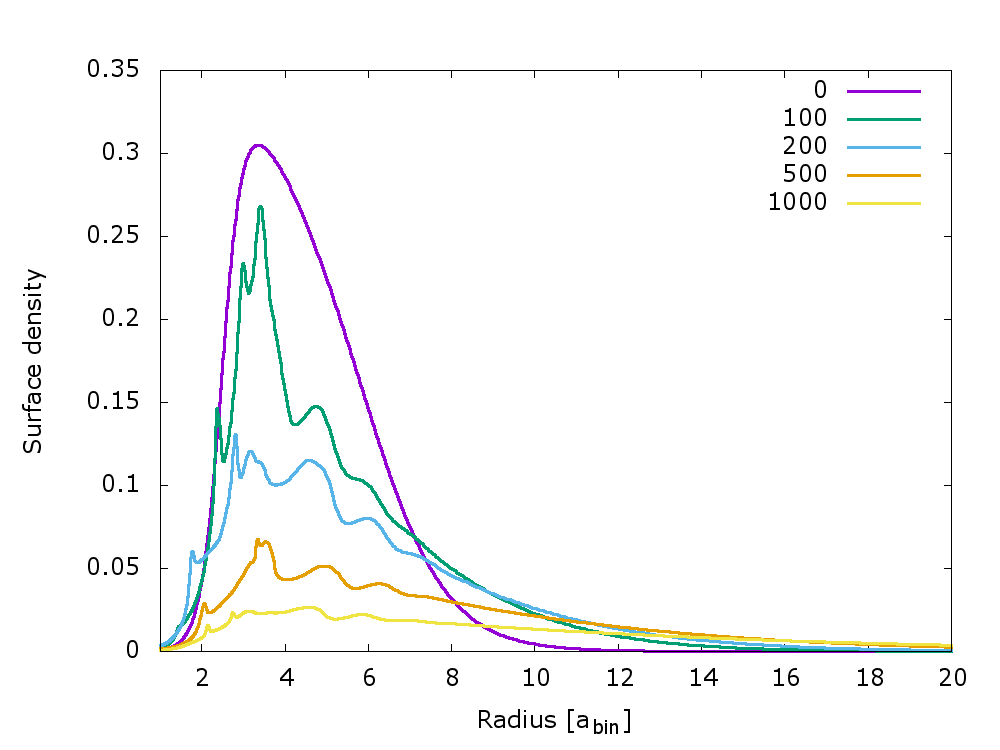}
    \caption{Azimuthally averaged surface density for the standard model ($q=0.5, h=0.1, \alpha=0.1$)
    at five different times, quoted in binary orbits.
   }
    \label{fig:sigma_evol_standard}
\end{figure}

\begin{figure}[htb]
    \centering
    \includegraphics[width=0.45\textwidth]{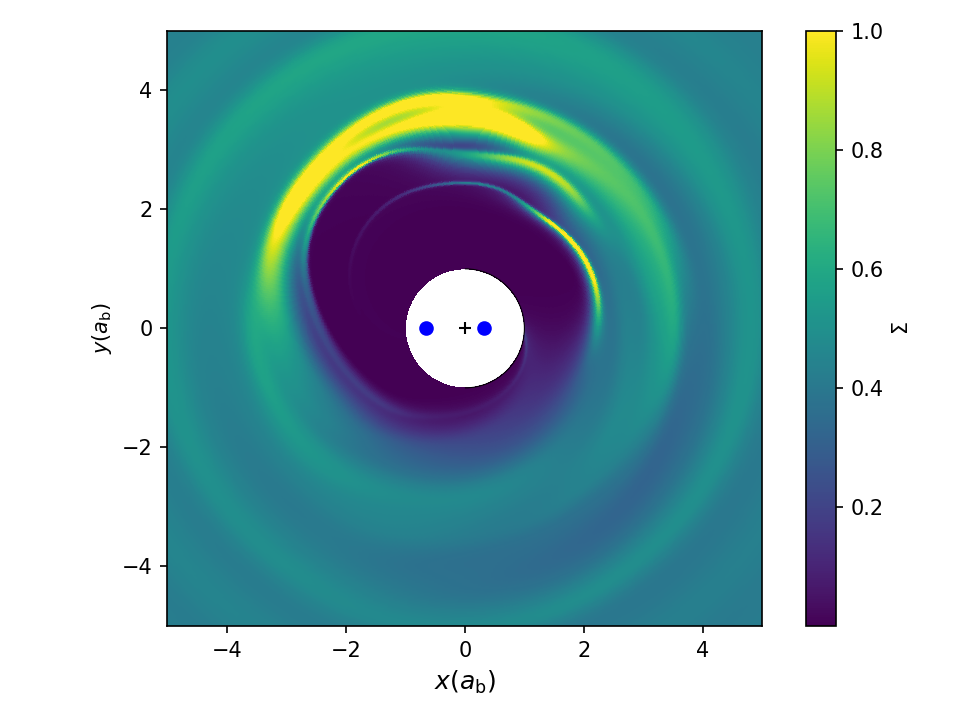}
    \caption{The 2D surface density distribution at $500\,T_\mathrm{bin}$ for the standard setup.
   The black cross marks the binary's barycenter and the region within
  $r=1\,a_\mathrm{bin}$ is not part of the computational domain.
    }
    \label{fig:2d}
\end{figure}

\begin{figure}[th]
    \centering
    \includegraphics[width=0.45\textwidth]{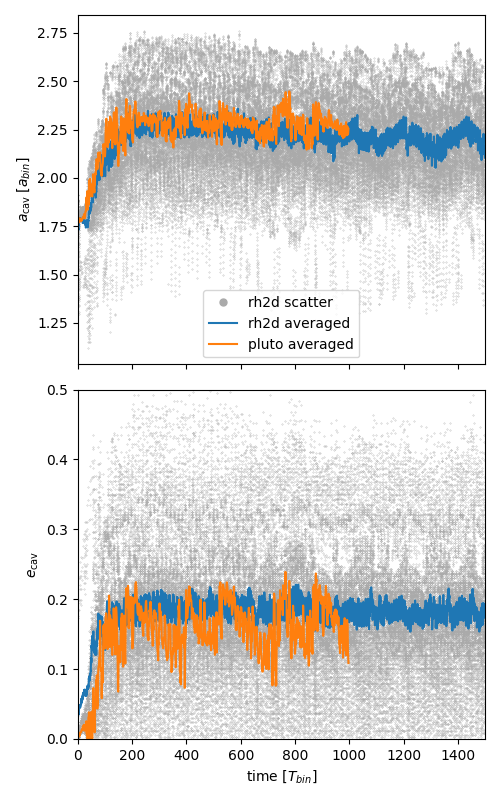}
    \caption{Approximated cavity size and eccentricity for the standard model.
     The curves denoted 'averaged' denote the sliding averages over 10 orbits
     for 3 different codes used. 
    The grey dots show the instantaneously estimated values.
   }
    \label{fig:ae_cav_standard}
\end{figure}

\begin{figure}[htb]
    \centering
    \includegraphics[width=0.45\textwidth]{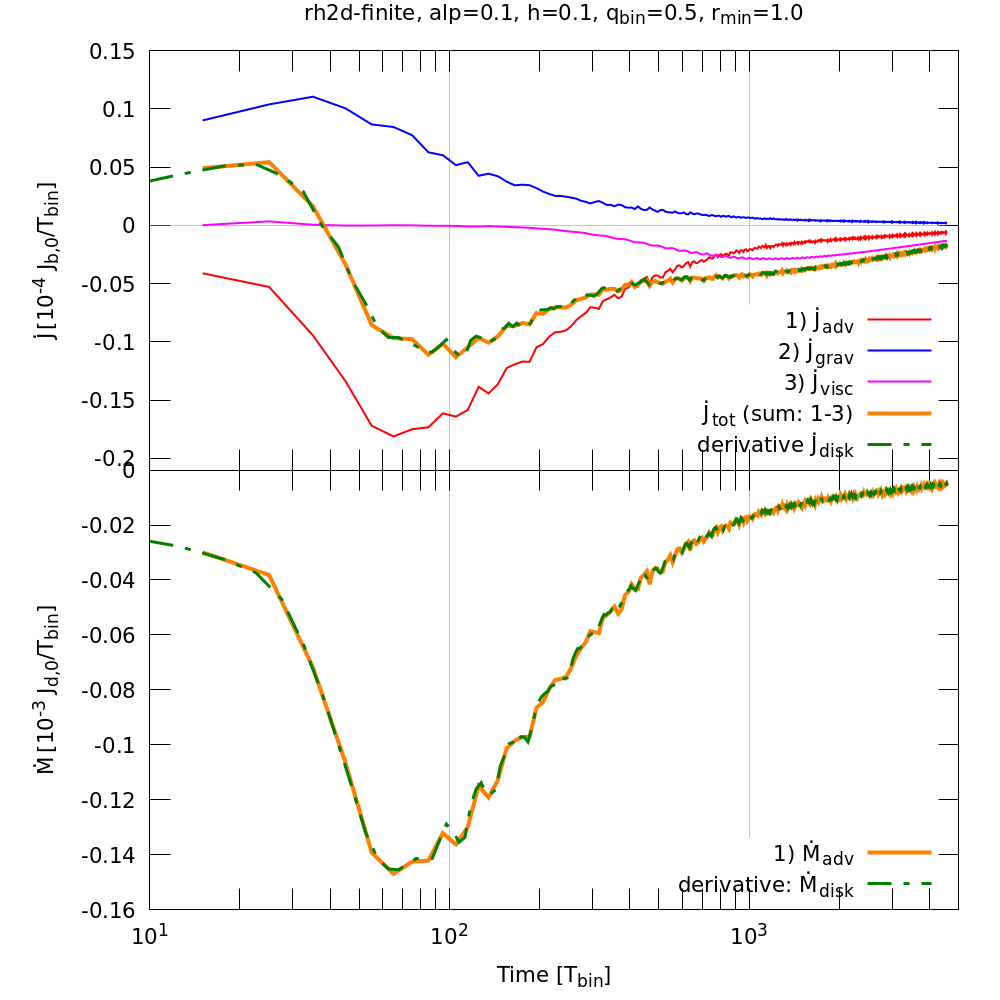}
    \caption{Loss/gain channels of mass and angular momentum (AM) of the disc.
     The curves denote: 1) $\dot{J}_\mathrm{adv}$ (through $r_\mathrm{min}$), 2) $\dot{J}_\mathrm{grav}$ (torque by binary),
     and 3) $\dot{J}_\mathrm{visc}$ (through $r_\mathrm{max}$).
     The dashed curves labelled 'derivative' denote the time derivative of the disc's total mass/AM.
     The sum of 1-3 (green curve) matches exactly the calculated changes.
   }
    \label{fig:fluxesc_standard}
\end{figure}

\section{The standard model}\label{sec:standard}
In this section we present our results for the standard model which has the parameter stated in Table~\ref{tab:standard}.
First, a general overview of the main disc physics will be given, followed by a discussion on the choice of the inner radius of the domain, $r_\mathrm{min}$,
and then the expected binary evolution. 

\subsection{Disc structure}
\label{subsec:disk}
Most of the results quoted in this subsection are obtained with the \texttt{RH2D} code, unless otherwise stated.
Given the inner and outer radii, $r_\mathrm{in}=2.5$ and $r_\mathrm{out}=6$ for the standard model, an initial density setup similar to \citet{2020ApJ...889..114M}
was generated as shown in Fig.~\ref{fig:sigma_evol_standard}. The density has its maximum near $r=3.5 a$ and drops smoothly inwards and
outwards.
The initial and subsequent evolution of the azimuthally averaged surface density, displayed at five different snapshots in Fig.~\ref{fig:sigma_evol_standard},
shows a radial spreading of this initial ring-like density, while the maximum of $\Sigma(r)$ remains always close to the initial one.
The evolution is very fast for this setup, already at $t = 200\,T_\mathrm{bin}$ the density maximum has dropped by about a factor of three,
and the disc has lost about $20\%$ of its initial mass through the inner boundary.
To give an impression of the dynamical features of the disc flow in the vicinity of the binary we present in Fig.~\ref{fig:2d} the 2D density
distribution after a simulated time of $500\,T_\mathrm{bin}$.

\begin{figure}[htb]
    \centering
    \includegraphics[width=0.45\textwidth]{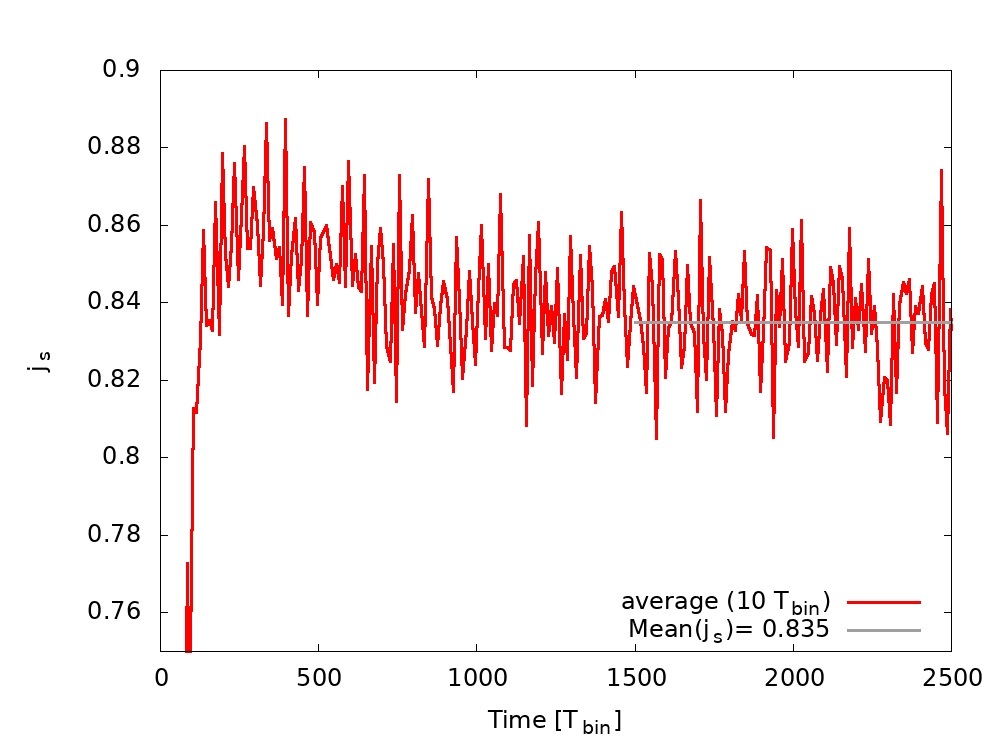}
    \caption{Specific angular momentum transfer, $j_\mathrm{s}$, onto the binary for the
     standard model. The curve represents a sliding average over 10 orbits, and the grey line shows the
     average over the last 1000 orbits.
   }
    \label{fig:js-standard}
\end{figure}

To have an indicator of the size and distortion of the inner cavity as the simulations evolves, we fit approximate ellipses to its shape using the
procedure described in \citet{2017Thun}. 
The result for the semi-major axis $a_\mathrm{cav}$ and eccentricity $e_\mathrm{cav}$ of the cavity is displayed in
Fig.~\ref{fig:ae_cav_standard} for the two different codes used in our main study.
After a brief initial growth phase, lasting for about 200-300 binary orbits, the cavity size reaches
a maximum radial extent of $a_\mathrm{cav} \sim 2.25a$ and a moderate eccentricity of $e_\mathrm{cav} \sim 0.18$.  
The shape of the cavity is (for these disc parameter) only an approximate ellipse,
such that the automatic fitting routine \citep[see][]{2017Thun} may be misguided by local density maxima peaks as seen in Fig.\,\ref{fig:2d}.
As a consequence the values vary a lot as indicated by the grey shaded points.
This issue is significantly reduced for simulations with smaller viscosities as the density inside the spirals is reduced and the peak density is clearly located at the apocentre of the inner disc, see also \citet{2021A&A...645A..68P}.
The results for the two grid-codes agree reasonably well. Differences can be due the fact that in \texttt{RH2D} results are computed automatically 'on the fly'
while the simulations are running, whereas for \texttt{PLUTO} they are obtained in a post-processing step from different snap-shots, here once every $T_\mathrm{bin}$.
In the following we will display smoothed results, here averaged over $10\,T_\mathrm{bin}$.

\subsection{Mass and angular momentum balance}
During the computation we monitor the global mass and angular momentum (AM), and
the fluxes across the outer and inner boundary as well.
To obtain an idea of the importance of the individual contributions, we display the individual loss/gain channels in Fig.~\ref{fig:fluxesc_standard}, labelled by 1) to 3).
In addition, we display the sum of
all 3 contributions and show that it matches exactly the rate of change in mass and AM as calculated from the time evolution
of the global disc values, denoted by 'derivative' in the plot.
This shows that our measurements are fully consistent and that we can accurately track all loss/gain channels of mass and AM.
As seen in Fig.~\ref{fig:fluxesc_standard} the torques from the binary create a positive input of AM to the disc (blue curve) while the
loss through the inner boundary represents a loss term (red curve). The sum (in green) shows that during a short initial phase ($\approx 40\,T_\mathrm{bin}$)
the disc gains more angular momentum from the binary than it loses. After that the losses overwhelm and the disc loses angular momentum to the binary.
The loss of angular momentum through the outer boundary (purple curve) is negligible for the first $1000\,T_\mathrm{bin}$.
However, it begins to be the main loss channel after about $2300\,T_\mathrm{bin}$ when the disc has spread so much that
$| \dot{J}_\mathrm{visc}| > | \dot{J}_\mathrm{adv} |$.

The results on the mass and AM fluxes displayed in Fig.~\ref{fig:fluxesc_standard} are then used to calculate the specific angular momentum transfer,
$j_\mathrm{s}$, to the binary according to eq.~(\ref{eq:js-normalising}).
The outcome is displayed in Fig.~\ref{fig:js-standard}, where we plot $j_\mathrm{s}$ as a function of time for the standard model.
Initially,  $j_\mathrm{s}$ rises steeply until is reaches a maximum at $t \approx 300\, T_\mathrm{bin}$, after that it drops slowly to reach an equilibrium
value after about $1000\, T_\mathrm{bin}$ of about $j_\mathrm{s} = 0.84$. This equilibrium of $j_\mathrm{s}$ is obtained despite the fact
that the overall disc density, mass and AM are still evolving in time, and formally not a steady state has been reached. The reason lies in the
fact that the inner disc region has reached dynamical equilibrium as Fig.~\ref{fig:ae_cav_standard} already indicated. The angular momentum flux
$j_\mathrm{s}$ represents a eigenvalue of the disc as shown already by \citet{2017Miranda} and \citet{2019Munoz}.

Using the measured $j_\mathrm{s} = 0.84$ and the quoted critical values in Fig.~\ref{fig:ana_js}, we can argue that
this will lead to an expansion of the binary's orbit for a mass ratio $q=0.5$ and the chosen disc parameter because
$j_\mathrm{s} = 0.84 > \jscrit$ for all possible accretion ratios $f$.
This statement about the orbital evolution of a circular binary is possible even though we do not include the binary explicitly in the computational domain,
because all the mass and angular momentum that crosses the inner boundary has to be accreted by the binary.
A necessary requirement is the establishment of a quasi-stationary state at least in the inner regions of the disc which can be checked
for example by monitoring properties such as $a_\mathrm{cav}$, $e_\mathrm{cav}$, and of course $j_\mathrm{s}$ directly.
The value of $r_\mathrm{min}$ must be chosen small enough to include the relevant disc dynamics and the diode inner boundary condition does not impact the results.
In the next section we show that the measured value of $j_\mathrm{s}$ does not depend on the exact location of the inner boundary once $r_\mathrm{min} \lesssim a_\mathrm{bin}$.

\begin{figure}[t]
    \centering
    \includegraphics[width=0.45\textwidth]{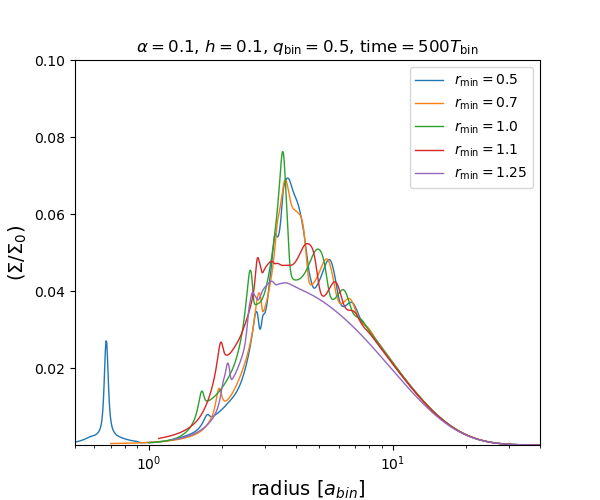}
    \caption{Azimuthally averaged density for five different inner grid radii $r_\mathrm{min}$
    after 500 binary orbits, for simulations using the \texttt{PLUTO} code.
   }
    \label{fig:sigc_500_rmin}
\end{figure}

\begin{figure}[htb]
    \centering
    \includegraphics[width=0.45\textwidth]{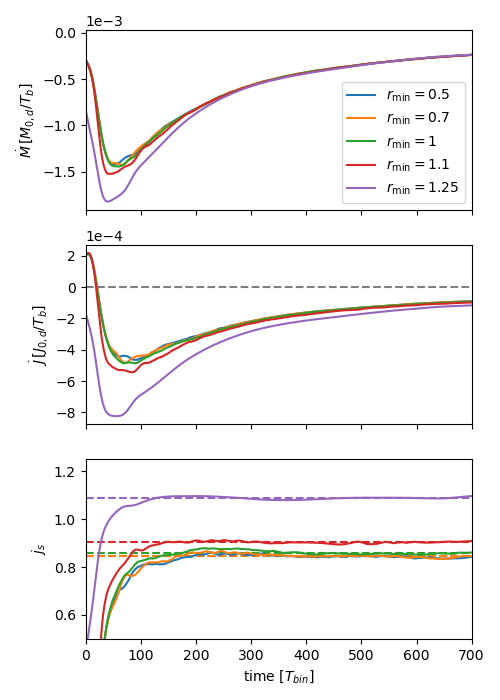}
    \caption{Evolution of mass and angular momentum gain/loss rates of the disc, and $j_\mathrm{s}$ for different inner disc radii,
    using the \texttt{PLUTO} code. The dashed lines in the bottom panel indicate the obtained average values between $200-700\,T_\mathrm{bin}$.
   }
    \label{fig:flux_rmin}
\end{figure}

\subsection{Varying the inner radius of the grid}
\label{subsec:rmin}
The determination of the angular momentum flux $j_\mathrm{s}$ depends on the details of the disc structure
around the binary. Knowing from previous studies \citep{2017Thun} this can depend on the location of the inner grid radius, $r_\mathrm{min}$.
In order to validate our results and check for the numerical impact of this choice we performed additional computations
using different values $r_\mathrm{min} =$ 1.25, 1.1, 1.0, 0.7 and 0.5. For this study we adapted the number of radial grid cells such that
we always had the same spatial resolution in the overlapping region.
In Fig.~\ref{fig:sigc_500_rmin} we display the azimuthally averaged surface density at 500 binary orbits
for the standard model for the different $r_\mathrm{min}$, now for the whole spatial domain.
In the outer disc region, beyond roughly $4\,a$ the density distributions are very similar for all $r_\mathrm{min} \leq 1.0\,a$.
They all reach similar maximum densities, however, the two models with larger $r_\mathrm{min}$ (1.1, 1.25) have substantially smaller peak values,
and are much smoother.
This means that in spirals in the disc are reduced that cause perturbations in the density profile. The $r_\mathrm{min}=1.25$ model displays not more spirals, and is therefore dynamically very different.
For the model with the smallest $r_\mathrm{min}=0.5$ there is a pronounced maximum in the density distribution at the location of the secondary star
because material is collecting in its potential minimum.
Concerning the overall size and shape of the cavity, i.e. its eccentricity and semi-major axis, we also find good agreement for the smaller $r_\mathrm{min}$,
and deviations for the two larger values.

A similar result is obtained for the important mass and AM balance of the disc. In the top panels of Fig.~\ref{fig:flux_rmin} we display the time
evolution of the disc's mass and AM loss/gain rates. For the models with $r_\mathrm{min} \lesssim 1$ 
they are identical during the whole evolution, however for larger inner boundaries the accretion rates increases non-physically.
For the specific angular momentum accretion $j_\mathrm{s}$ onto the binary,
displayed in bottom panel of Fig.\,\ref{fig:flux_rmin},
the results are very close as well for the three models with the smaller $r_\mathrm{min}$.
Hence, for sufficiently small $r_\mathrm{min}$ the evolution is indeed solely given by the physical parameter of the system.
Thereby, we conclude that $r_\mathrm{min}=1.0$ is sufficient even for the highest values of viscosity and aspect ratio where the inner region
around the two stars is not entirely empty. Hence, we adopt
this value for our subsequent models and the parameter study as it represents a good compromise between physical accuracy and computational speed.

\begin{figure}[t]
    \centering
    \includegraphics[width=0.45\textwidth]{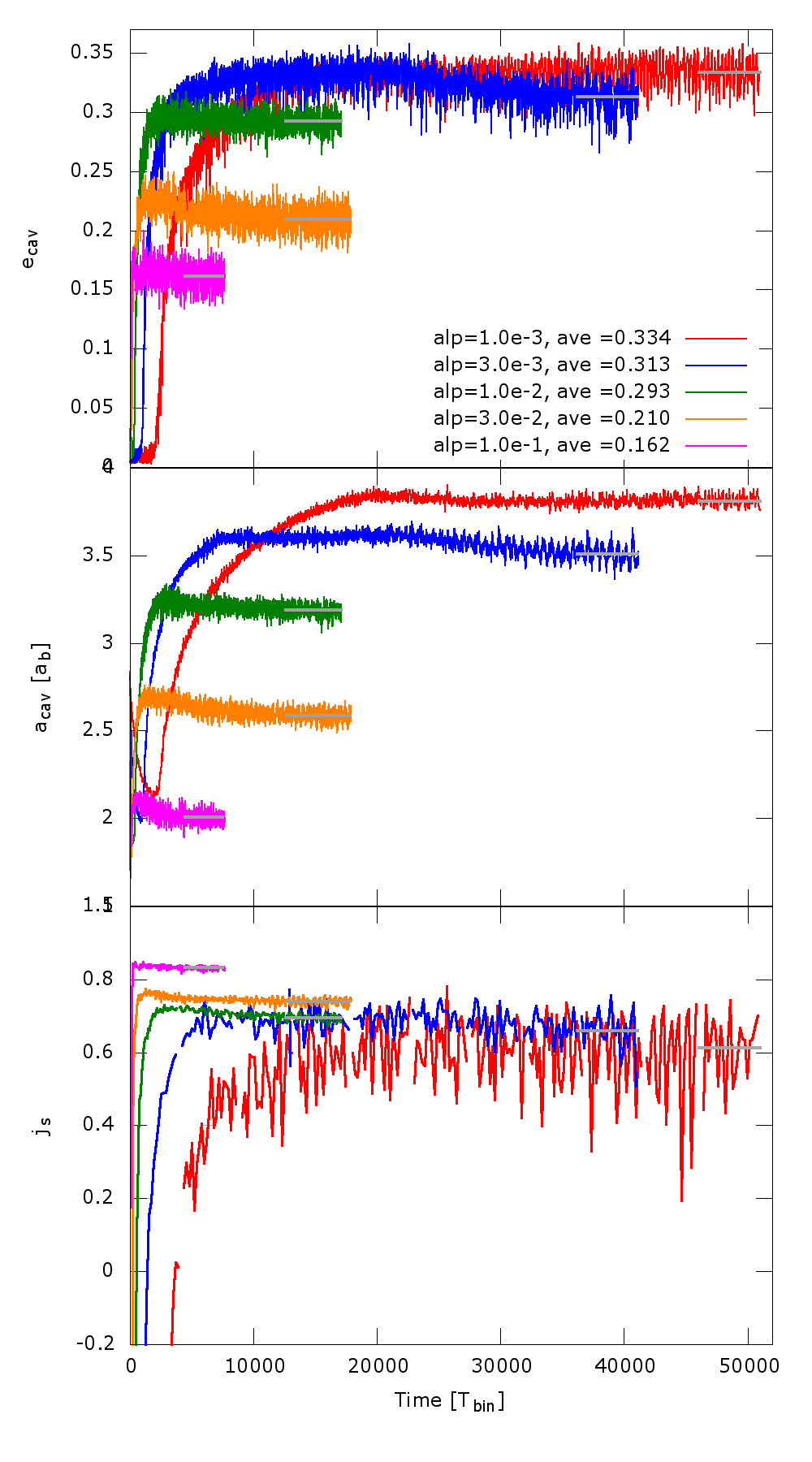}
    \caption{The shape and size of the cavity around the binary (with $q=0.5$) and the specific angular momentum $j_\mathrm{s}$
    transferred, as a function of time for models with different viscosity, and an aspect ratio $h=0.1$.
    The top two panels show sliding time averages over $10\,\Tbin$, in the bottom panel
    it was $200\,\Tbin$ for the two lowest $\alpha$ models, over $100\,\Tbin$
   for $\alpha=0.01$, and over 50 $\Tbin$ for $\alpha=0.03$ and $0.1$.
   The grey lines indicate the averaging interval for the final quoted equilibrium values.
   Simulations were performed with the \texttt{RH2D}-code.
    }
    \label{fig:aecavjs-alp}
\end{figure}

\section{Parameter studies}\label{sec:parameters}
After having analysed the physics of the standard model we investigated in more detail the dependence on the disc physics
and ran models for different viscosities and aspect ratios.
Additionally, we varied the mass ratio of the binary.

\subsection{Varying $\alpha$ for the standard model}
\label{sec:alpha}

\begin{figure}[htb]
    \centering
    \includegraphics[width=0.45\textwidth]{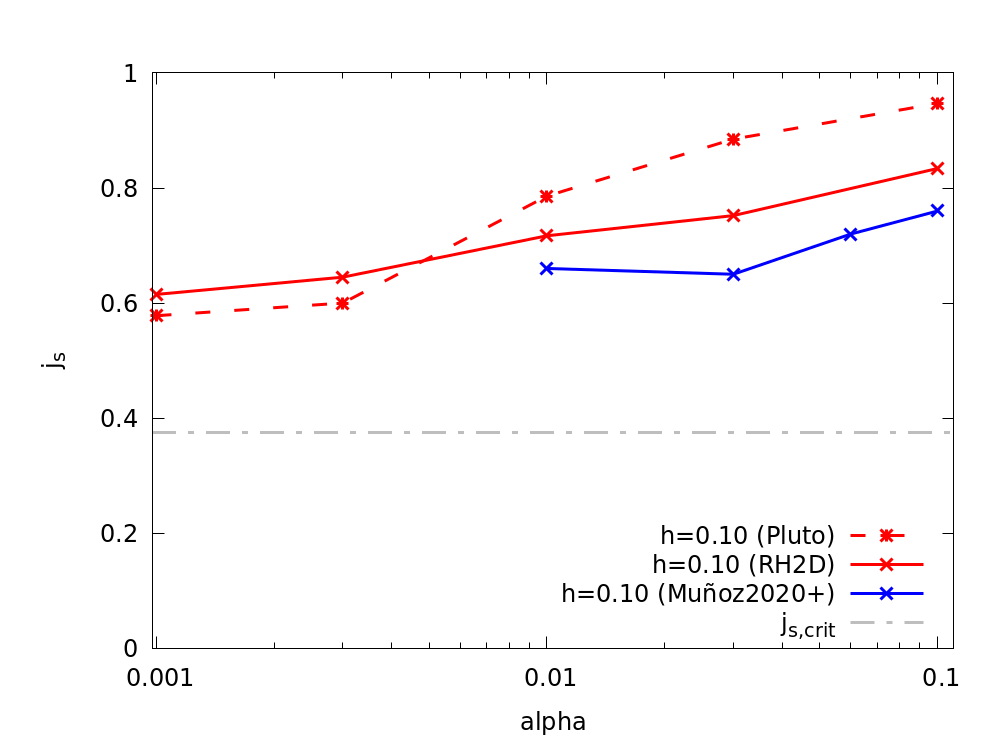}
    \caption{The specific angular momentum $j_\mathrm{s}$ transferred to the binary (with $q=0.5$)
    as a function of the viscosity with an aspect ratio of $h=0.1$.
    The red data points and lines refer to the models in this work using two different codes, and the blue points and line refer to the results of \cite{2020ApJ...889..114M}.
    }
    \label{fig:jsc-munoz}
\end{figure}

To study the impact of viscosity 
we present here models using a different $\alpha$-parameter with values of [0.1,0.03,0.01,0.003,0.001],
while keeping $H/r=0.1$ and $q=0.5$.
This allows a direct comparison to \citet{2020ApJ...889..114M} who performed a similar study using four different $\alpha$ values.
For lower viscosities the expected cavity size will become larger because the viscous torques are reduced \citep{1994ApJ...421..651A,2015MNRAS.452.2396M}.
To account for this, we changed the initial profile and moved the ring further out by choosing $r_\mathrm{in}=4\,a$ and $r_\mathrm{out}=10\,a$ in eqs.\,((\ref{eqn:g_1}) and (\ref{eqn:g_2}).
We have checked that identical equilibria, concerning the dynamics of the inner disc region (i.e. $a_\mathrm{cav}, e_\mathrm{cav}$ and $j_\mathrm{s}$), 
are obtained independent of the choice of the initial density distribution.

To give an impression of the evolution towards equilibrium we show in Fig.~\ref{fig:aecavjs-alp} the time evolution of the cavity size, its eccentricity 
and $j_\mathrm{s}$ as a function of
time for the 5 different $\alpha$-values. 
Again, we obtain $a_\mathrm{cav}$ and $e_\mathrm{cav}$ using the fitting procedure described in
\cite{2017Thun}.
As expected, the models with the lower viscosities take much longer to reach equilibrium as this is given by the viscous timescale.
The time needed for the $\alpha=10^{-3}$ model to fully evolve is $> 30\,000\,\Tbin$.
The $\alpha=0.1$ model is used most often in the literature, because the equilibration time is by far the shortest.
For this case the cavity size is smallest ($a_\mathrm{cav}=2.0$) with a mean eccentricity of $0.16$.
Lowering the viscosity, $a_\mathrm{cav}$ and $e_\mathrm{cav}$ increase monotonically and reach
a maximum size and eccentricity for the model with the smallest $\alpha=0.001$.

The corresponding evolution of the normalised angular momentum accretion onto the binary is displayed in the bottom panel of Fig.~\ref{fig:aecavjs-alp}.
To obtain the final $j_\mathrm{s}$ value we average over the length marked in grey and obtain the quoted values.
The averaging time increases for smaller viscosities due to larger noise in the data.
During the initial cavity filling phase $j_\mathrm{s}$ is negative because the gravitational torques overwhelm the 
inflow of angular momentum $|\dot{J}_\mathrm{grav}| > |\dot{J}_\mathrm{adv}|$ as seen for example in 
\citet{2020ApJ...889..114M} and \citet{2020ApJ...900...43T}.
Again, the approach to equilibrium proceeds on the same (long) viscous timescales, and the final values show a monotonic 
drop with viscosity.

In Figure \ref{fig:jsc-munoz} we display the obtained $j_s$ as a function of viscosity for the two different codes we used (\texttt{RH2D} and \texttt{PLUTO}).
Both series show very similar behaviour, an increase of the angular momentum transfer to the binary with increasing viscosity.
Overplotted are results extracted from \citet{2020ApJ...889..114M} which are for this study slightly lower than ours.
The deviations in $j_s$ could be caused by numerical differences like time steps or different grid structures as \cite{2020ApJ...889..114M} uses mesh refined grids but also be the result of the occurring uncertainty in the measurement of $j_s$. However, all models agree in the fundamental finding that thick discs with $h=0.1$ lead to expanding orbits, even for viscosities as low as $\alpha=10^{-3}$.

\subsection{Models with constant viscosity}
\label{sec:Mach}

\begin{figure}[t]
    \centering
    \includegraphics[width=0.45\textwidth]{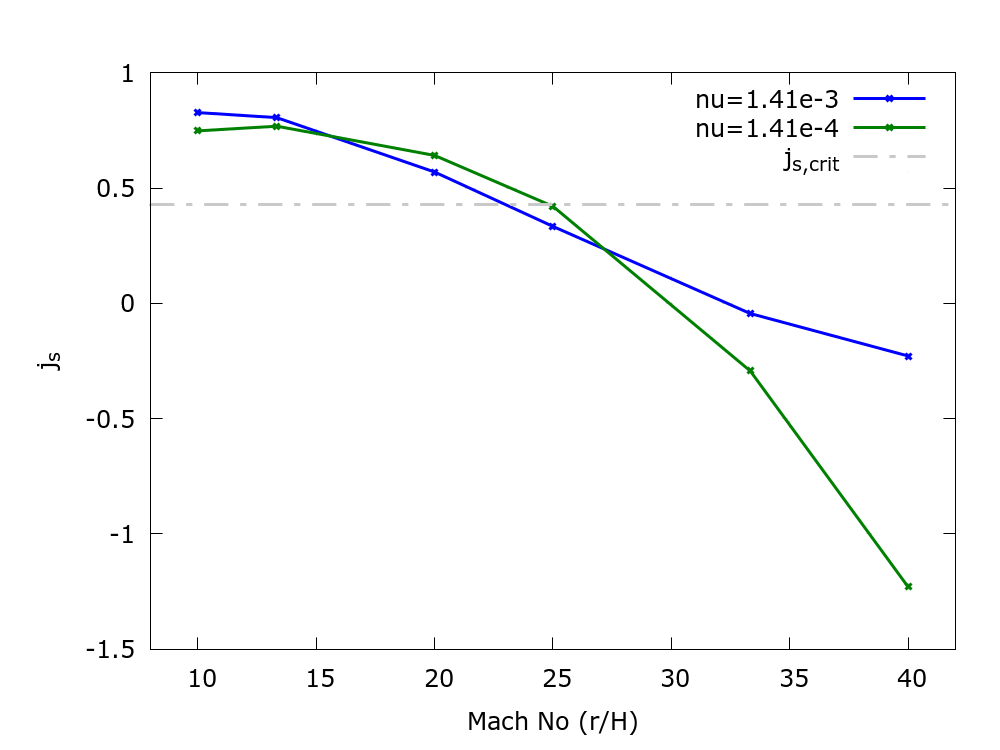}
    \caption{The specific angular momentum $j_\mathrm{s}$ transferred to the binary as a function of the disc's Mach number for models
    with two different values of the kinematic viscosity, for a mass ratio $q=0.5$.
    The dashed dotted line refers to a critical $\jscrit=0.43$.
    }
    \label{fig:const_qbin}
\end{figure}

\begin{figure}[t]
    \centering
    \includegraphics[width=0.45\textwidth]{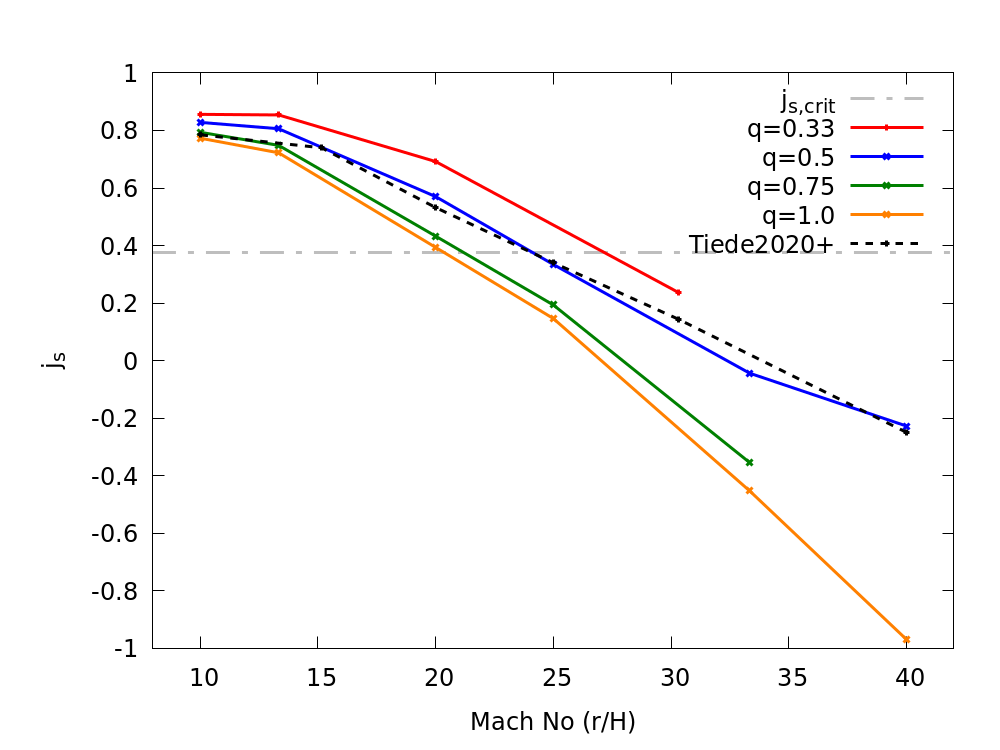}
    \caption{The specific angular momentum $j_\mathrm{s}$ transferred to the binary as a function of the disc's Mach number.
    Presented are models using a constant kinematic viscosity, $\nu = \sqrt{2} \times 10^{-3}$, for different values of
    the binary mass ratio. The dashed dotted line refers to the critical $\jscrit=3/8$ for the case $q=1$.}
    \label{fig:const_nu}
\end{figure}

To allow for an individual and independent variation of viscosity and disc scale height,
we ran a series of models using a constant kinematic viscosity $\nu$ and varied disc thickness and mass ratio.
We studied two different values of $\nu$, first we took the same value as \citet{2020ApJ...900...43T} $\nu = \sqrt{2} \times 10^{-3} \,[a^2\,\Omega_\mathrm{bin}]$,
which allows a direct comparison to their work, and then a ten times lower value.
The chosen $\nu$ refer to $\alpha$ values of 0.1 and 0.01 at a distance of $r = a$ for $H/r=0.1$.
In Fig.\,\ref{fig:const_qbin} $j_\mathrm{s}$ is displayed as a function of the disc's Mach number ($=r/H$) for a binary with our standard mass ratio $q=0.5$, 
for the two different viscosities. 
For both viscosity values we find that $j_\mathrm{s}$ drops with Mach number and turns negative for Mach numbers between 30 and 33.
This implies that in thin discs the binary orbit will expand where the turning point lies around a Mach number of 23-25, given by the point at which the curves cross the grey
dashed-dotted line denotes a $\jscrit = 0.43$, calculated for a mass accretion ratio of $f=1.6$. See appendix ~\ref{subsec:f} for a discussion about the mass accretion ratio.

In addition to their study we varied now the mass ratio (ranging from $q=0.33$ to $q=1.0$),
and display our results in Fig.~\ref{fig:const_nu}. 
We find the trend that smaller mass ratios $q$ lead to an increase in $j_\mathrm{s}$,
while the general behaviour of decreasing $j_\mathrm{s}$ with increasing Mach number is not impacted significantly by changing $q$.
To determine the transition point between expansion and shrinkage for the values of $q \neq 1$ requires knowledge of the mass accretion
factor $f$. For a rough estimate we added the dashed dotted line, which refers to $\jscrit=3/8$ for $q=1$. The true values for the other
$q$ will be only slightly larger as argued in appendix ~\ref{subsec:f}. 
For equal mass binaries we find an orbit expansion for thicker discs with Mach number smaller than $\sim 20$.
To compare to earlier results, we overplot the results of \citet{2020ApJ...900...43T} obtained for $q=1$ (black dashed line).
While the general trend is similar the absolute values differ, possibly due to the different numerical methods applied. Our simulations are 2D on a cylindrical grid
not containing the binary while theirs are 2D Cartesian.

\subsection{Parameter space of H/r and $\alpha$}
\label{subsec:h-alpha}
\begin{figure}[t]
    \centering
    \includegraphics[width=0.45\textwidth]{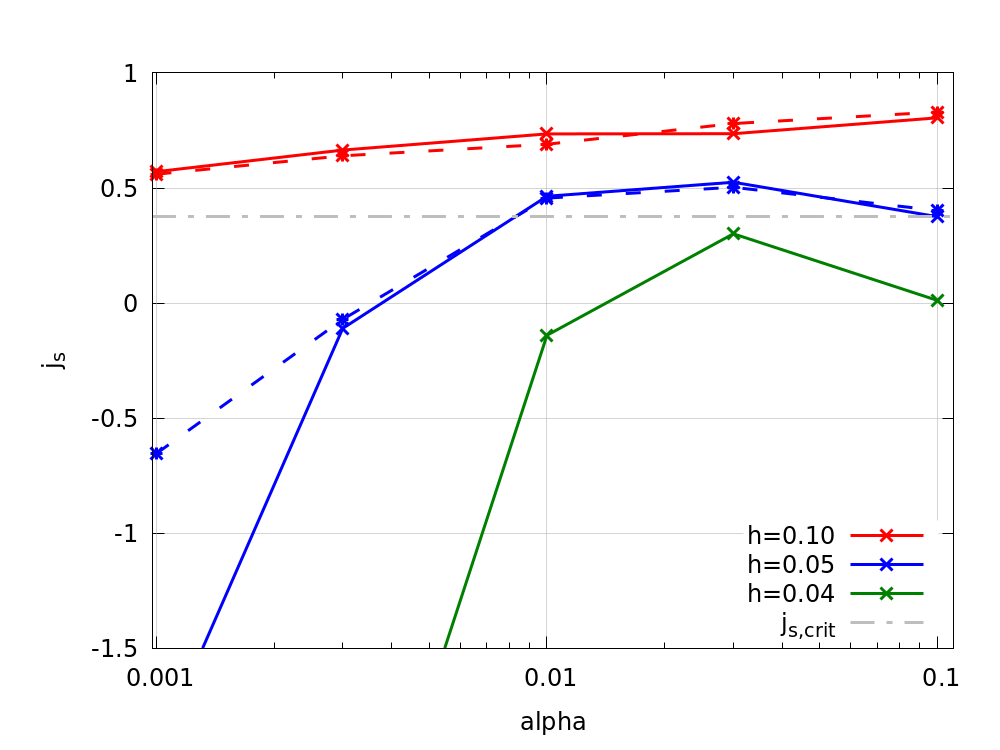}
    \caption{The specific angular momentum $j_\mathrm{s}$ transferred to the binary as a function of the disc's viscosity,
    for a mass ratio, $q=1.0$ and different aspect ratios.
    For $q=1$ the critical rate is given by $j_\mathrm{s,crit}$=3/8.
    Results for two different codes, \texttt{RH2D} (crosses, solid lines) and \texttt{PLUTO} (asterisks, dashed lines), are displayed.
    }
    \label{fig:vary_alp_q10}
\end{figure}

\begin{figure}[t]
    \centering
    \includegraphics[width=0.45\textwidth]{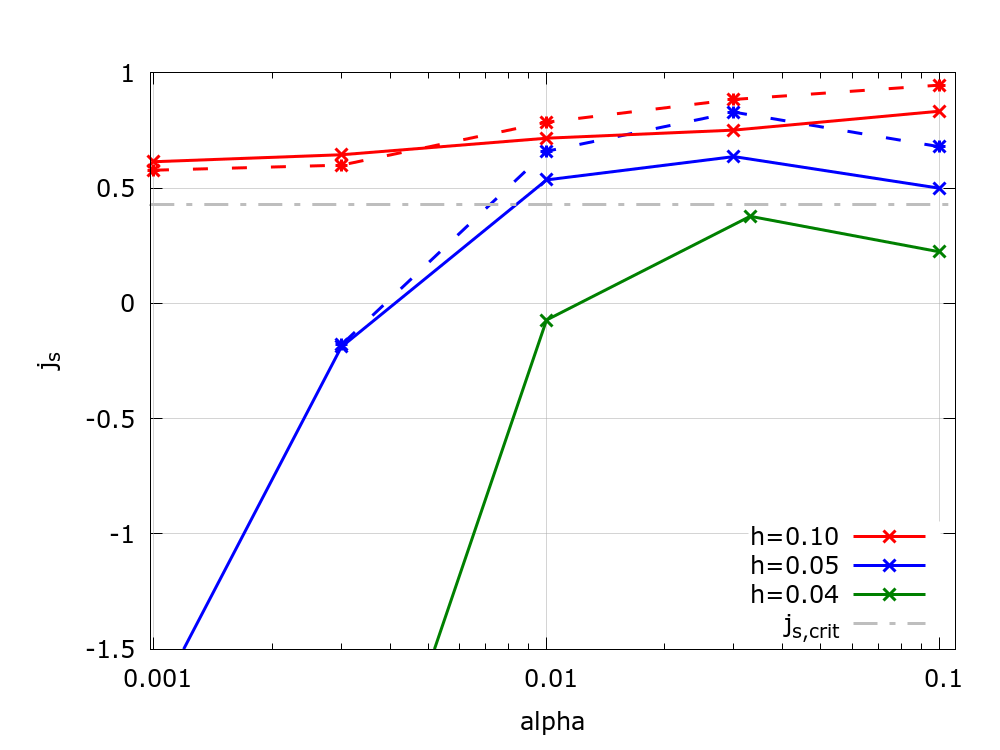}
    \caption{The specific angular momentum $j_\mathrm{s}$ transferred to the binary as a function of the disc's viscosity,
    for a mass ratio, $q=0.5$ and different aspect ratios, using the same notation as in Fig.~\ref{fig:vary_alp_q10}.
    For the critical rate we assumed here $q=0.5$ and $f=1.6$, and obtain $j_\mathrm{s,crit}$=0.43.
    }
    \label{fig:vary_alp_q05}
\end{figure}

Here, we extended our parameter study to standard $\alpha$-disc models models using five different values
$\alpha = [0.1, 0.03, 0.01, 0.003, 0.001]$ and 3 different aspect ratios $H/r=[0.1, 0.05, 0.04]$, both for mass ratios of $q=0.5$ and $q=1$.
As explained in Sect.\,\ref{sec:alpha} we run all the models long enough to reach a quasi-stationary equilibrium state with constant $a_\mathrm{cav}$,
$e_\mathrm{cav}$ and $\js$.

In Fig.~\ref{fig:vary_alp_q10} the results for $q=1$ are displayed obtained with two different codes, \texttt{PLUTO} and \texttt{RH2D}. The results obtained with \texttt{RH2D} are shown
by the crosses and solid lines while the \texttt{PLUTO} results feature the asterisks and the dashed lines.
The grey dashed-dotted line denotes the critical $\jscrit = 0.375$ for $q=1$. As already shown in Sect.~\ref{sec:Mach} above, for a thick disc with $h=0.1$
expanding binary orbits are possible for all values of the viscosity (at least down to $\alpha=0.001$).
This is due to the fact that in thick discs the high pressure leads to a continuous
refill of the inner cavity at a high rate carrying a significant amount of angular momentum with it.
For thinner discs the pressure effect is reduced as well as the viscosity, which scales with $h^2$, and the resulting $j_\mathrm{s}$-values become smaller,
such that the possibility of expanding orbits will be reduced.
The smallest aspect ratio, $h=0.04$, results only in shrinking binary orbits, similar to what has been
discussed already in section~\ref{sec:Mach} above for the case of constant viscosities.
For $h=0.05$ only for a small window of viscosities between $\alpha = 0.01$ and 0.1 an expanding orbit is possible.
We find a maximum in $j_\mathrm{s}$ for $\alpha\approx0.03$ and a steep decrease of $j_\mathrm{s}$ towards smaller viscosities.

In systems with a mass ratio of $q=0.5$, as shown in Fig.~\ref{fig:vary_alp_q05}, the same trends apply. The denoted $\jscrit$ corresponds to an accretion ratio
$f=1.6$, as motivated by our accretion studies presented in section \ref{subsec:f}. The overall shape of the curves is very similar to the $q=1$ case.
For large viscosities $\alpha \ge 0.01$ the values for $j_\mathrm{s}$ lie above the $q=1$ case, but as $\jscrit$ is also increased,
the possible parameter range in which orbit expansion is possible remains the same as in the case of equal mass binaries.

For our two model sequences, $q=1.0$ and $0.5$, shown in Figs.~\ref{fig:vary_alp_q10} and \ref{fig:vary_alp_q05},
we performed for the thicker discs with $h=0.1$ and 0.05 comparison simulations with our two grid codes (\texttt{RH2D}, \texttt{PLUTO}).
In general the agreement between the two codes is very good. Both show an identical behaviour upon changing the disc viscosity and scale height,
even though the absolute values show some differences. Only for the lowest viscosity, $\alpha = 10^{-3}$, the differences become more pronounced but these models
suffer more from numerical effects such as density floor, or diffusion.

In the analysis for very low viscosities the uncertainties in the measured and averaged $j_\mathrm{s}$ increase, and a much longer
time evolution of several $\geq10\,000\,T_\mathrm{bin}$ is needed to reach a stable state of the disc.
Additionally, the finite density floor begins to affect the results due to the very low density in the cavity.
Even given these challenges in the simulations, the general picture emerges that low viscosities and small aspect ratios
lead to shrinking binary orbits as the cavities become very deep and the advected angular momentum decreases.

\begin{figure}[t]
    \centering
    \includegraphics[width=0.45\textwidth]{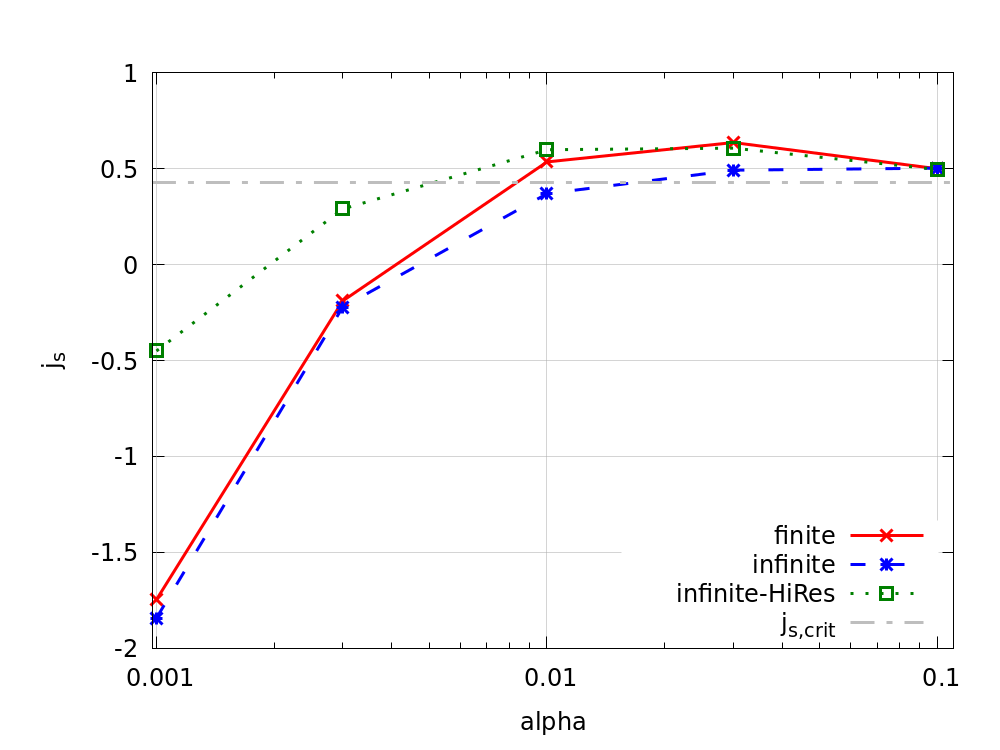}
    \caption{Comparison of the normalised angular momentum accretion, $j_\mathrm{s}$, for model with $q=0.5$ and $H/R=0.05$ 
    for torus-like (finite) and infinite disc models using the standard (684x584) vs. higher (968x824) grid resolutions.
   }
    \label{fig:res_js}
\end{figure}

\subsection{Infinite disc models}\label{subsec:infinite}
The models discussed so far suffer from the problem that no genuine global stationary state is reached. 
To investigate a different physical setup and check for possible shortcomings of this approach we ran additional models using an 'infinite' disc setup,
where we covered the full viscosity range for the $q=0.5, H/R=0.05$ model in different numerical resolutions.
To model an infinite disc, $r_\mathrm{out}$ in eq.~(\ref{eqn:g_2}) is set to very large values, implying that $g_\mathrm{out} \rightarrow 1$.
At the same time we apply damping boundary conditions such that the density at $r_{max}=40$ is always damped towards the initial state \citep{2006MNRAS.370..529D}, while
the velocities are kept at their original values.
This setup will allow a quasi-stationary equilibrium state to be reached and is hence comparable to the infinite disc models of \citet{2020ApJ...889..114M}.
Due to the outflow condition at the inner boundary, at $r_{min}=a$, there is a mass flow through the disc and the (time averaged) mass accretion will be radially constant, see also
\citet{2014Kley}.
It is crucial to set a very low density floor in the models, because the small viscosity simulations can reach very low densities inside the cavity.
Here, we used floor values as low as $10^{-10}$ of the initial $\Sigma(r=1)$.

The resulting $j_s$ is shown in Fig.~\ref{fig:res_js} where we display results for the finite torus models from above (red data termed 'finite')
and new simulations using the 'infinite' disc setup at two different grid resolutions, the standard 684x584 (blue-dashed) and a higher resolution 968x824 case (green-dotted).
To save computational time, the higher resolution models are not started from $t=0$ but are continued from the standard resolution runs at a time when those reached
approximately their equilibrium state. 
All simulations in Fig.~\ref{fig:res_js} follow the same overall trends but differ in the actual values of $j_\mathrm{s}$.
Concerning $j_\mathrm{s}$, we find an increase for the higher resolution runs for the lower viscosities, however it is not sufficient to lead to an expansion of the binary because the values lie consistently below $\jscrit$.
We confirm the finding of \cite{2020ApJ...889..114M}, that the outer boundary will not significantly change the inner angular momentum transport relevant for the binary. 
Additional tests for $H/R=0.05, \alpha=0.003$ with even higher resolution show no further changes in $j_\mathrm{s}$, and
we believe the shrinking of the binary to be real for very low viscosities.

\begin{figure}[t]
    \centering
    \includegraphics[width=0.45\textwidth]{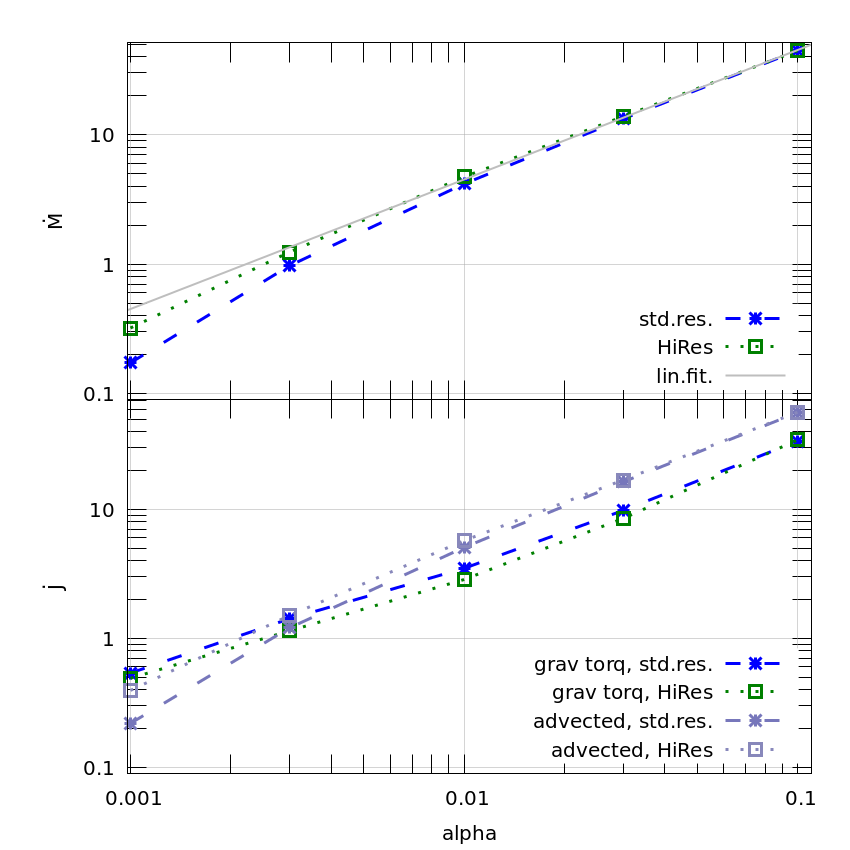}
    \caption{Top: The mass accretion rate onto the binary as a function of viscosity for infinite disc models with $q=0.5$ and $h=0.05$.
   The grey line indicates a linear slope.
   Bottom: The corresponding absolute value of gravitational (blue and green lines) and advective (grey lines) angular momentum transfer,
   $\dot{J}_\mathrm{grav}$ and $\dot{J}_\mathrm{adv}$ respectively.
    Shown are the results for the standard (asterisks, dashed lines) and higher resolution (squares, dotted).
   }
    \label{fig:res_mdotarat}
\end{figure}

For our locally isothermal infinite disc models in quasi-equilibrium for which the density at the outer boundary is fixed, the mass accretion rate should scale linearly with the viscosity, 
due to the classic relation $\dot{M} = 3 \pi \nu \Sigma$. In the top panel of Fig.~\ref{fig:res_mdotarat} we plot the mass accretion rate onto the binary (measured at $r_{min}$) as a function
of $\alpha$ for models with $q=0.5$ and $h=0.05$ at two numerical resolutions, using the line style of Fig.~\ref{fig:res_js}. 
Apart from the smallest $\alpha$-values the expected linear relation (grey line) is well satisfied, but
the agreement with the linear relation improves with higher grid resolution.
The bottom panel of Fig.~\ref{fig:res_mdotarat} shows the absolute value of the two contributions to $\dot{J}_\mathrm{b}$,
the gravitational part ($\dot{J}_\mathrm{grav}$, green and blue lines) and the
 advective part ($\dot{J}_\mathrm{adv}$, in grey). The gravitational term removes angular momentum from the binary, while the advection term adds angular momentum to it.
The two advective terms, $\dot{J}_\mathrm{adv}$ and $\dot{M}_\mathrm{adv}$, follow the same slope such that
the advective $j_\mathrm{s,adv}=\dot{J}_\mathrm{adv}/\dot{M}$ is almost constant at a value of $1.25\pm0.04$, upon varying the viscosity. 
It has a lowest value of $\sim 1.21$ for $\alpha =0.003$ and increases to $\sim 1.27$ for $\alpha=0.1$, and is $\sim 1.23$ for the smallest $\alpha=0.001$. 
At the same time the gravitational contribution $\dot{J}_\mathrm{grav}$ flattens towards smaller viscosities and equals the advective part for $\alpha \sim 0.002$, making the total $\js$ negative, as seen in Fig.~\ref{fig:res_js}.

\begin{figure*}[t]
    \centering
    \includegraphics[width=1.0\textwidth]{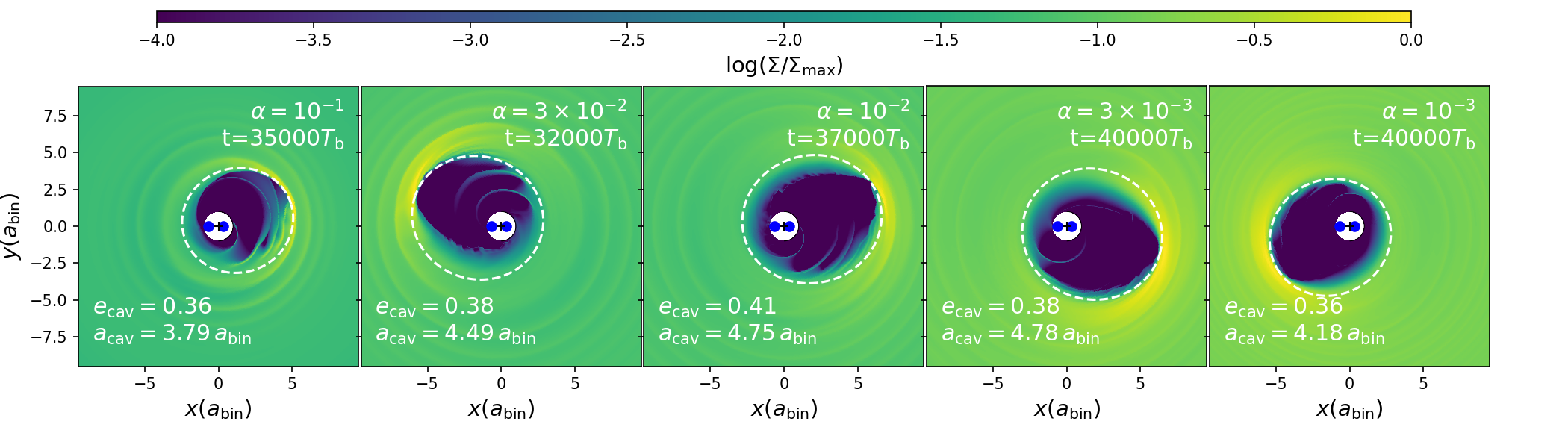}
    \caption{Snapshots of the surface density for infinite disc models with $q=0.5$ and $H/R=0.05$ for different viscosities, ranging from $\alpha =0.1$ (left) to
    $\alpha =0.001$ (right). Shown is the inner region of the disc for the higher resolution models in their final state at the times indicated.
    The white dashed lines indicate the ellipses fitted to the cavity with the parameters quoted. The blue dots mark the position of the stars. 
   }
    \label{fig:alpha-discs}
\end{figure*}

In order to understand the observed behaviour better we show in Fig.~\ref{fig:alpha-discs} the two-dimensional surface density distribution for the 'infinite' disc
models for the different viscosities.
The computed cavity edges are marked by the white, dashed ellipses, computed with the method described in \citet{2017Thun}.
Clearly seen are the eccentric inner cavities, which are more pronounced in these thinner discs ($h=5\%$) compared to the standard model (with $h=10\%$) even at a logarithmic scale.
The cavity's size and eccentricity are both smallest for the highest viscosity case ($\alpha=10^{-1}$) and increase upon lowering $\alpha$. Maximum values for $a_\mathrm{cav}$ and $e_\mathrm{cav}$ are reached for intermediate $\alpha$ and they drop again for even smaller viscosities.
Using the logarithmic colour scale the streamers and infalling gas inside the cavity are visible.
The strength of streamers weakens with reduced viscosity, 
coinciding with the generally declining trend in the advective angular momentum contribution seen in Fig.~\ref{fig:res_mdotarat}.
As expected the cavity becomes wider and deeper upon lowering the viscosity. When the viscosity becomes very low, however, this trend is slowed and even reverses.
At the same time with reduced accretion the mass in the streamers declines and torque from inside the cavity reduces proportionally.
However, the always eccentric disc causes the gravitational contribution in Fig.\,\ref{fig:res_mdotarat}
to change its slope around $\alpha=0.01$, aided by the slight contraction of the disc, resulting in shrinking binary orbits for the lowest viscosities.

\section{Discussion and conclusion}\label{sec:discussion}

We presented 2D hydrodynamical simulations of discs around a circular binary where disc and binary are coplanar. We monitored
the mass and angular momentum balance of the disc in order to calculate the binary's orbital evolution.
Our prime interest was to evaluate shrinkage or expansion of the binary 
by determining the sign of the semi-major axis evolution, for different
mass ratios of the binary and various disc parameter.
Based on the assumption that the binary's orbit remains circular under this accretion process we showed that its change in semi-major axis
due to the mass and angular momentum transfer between circumbinary disc and binary is directly proportional to the mass accretion rate times a
function that depends on the stellar mass ratio $q$, the specific angular momentum transfer $j_\mathrm{s}$, and the relative mass accretion onto
the binary $f$, see eq.~(\ref{eqn:adot_bin3}).

In agreement with the studies of \citet{2017Miranda} our simulations show that the specific angular momentum transfer rate $j_\mathrm{s}$ serves as an eigenvalue of the problem
that can be calculated using a cylindrical grid with an inner hole which has a radius of $r_\mathrm{in} \sim a_\mathrm{bin}$, i.e. the grid does not contain the central binary system.
Additionally, we confirm results of \cite{2020ApJ...889..114M} that in order to obtain a quasi-equilibrium for $j_\mathrm{s}$ it is sufficient to model a torus like configuration with limited radial extent,
because it is sufficient to bring the inner disc regions into equilibrium and not the whole disc.
This finding is supported by additional simulations using
an infinite disc setup that allows for a global stationary state with resulted in $\js$-values very similar to the torus models.

While for q=1 the critical value is given by $\jscrit=3/8$, for all $q \neq1$ knowledge of $f$ is required to find the exact threshold $0<\jscrit<1$. In the work of \cite{2020ApJ...889..114M} the mass accretion ratio of systems with $q=0.5$ averages $f\approx1.5$, which is consistent with the trend in \cite{2021ApJ...921...71D}.
In order to compare, we performed supplemental simulations covering the whole domain in the appendix  using Cartesian grids and the SPH method.
We found that for a mass ratio of $q=0.5$ the mass accretion ratio lies somewhere between 1.3 and 1.8, i.e.\ the secondary star accretes more mass from the disc
than the primary. As these studies are very time consuming and depend on numerical treatments such as potential smoothing, sink radii and others,
we will leave a more detailed analyses to future studies for the full range of parameters.

Using an average value $f = 1.6$ we showed for two mass ratios ($q=1$ and $0.5$) that binaries expand for sufficiently high viscosity, $\alpha \geq 0.01$, and disc
aspect ratios, $h \geq 0.05$. For thinner and less viscous discs the binaries tend to shrink.
Thick discs with $h=0.1$ result in expansion for all viscosities, while discs with $h=0.05$ show expansion for $\alpha$ larger than $\sim 0.005$.
Independent of the mass ratio we find that thin and low viscosity disc cause the binary orbit to shrink,
only thick and highly viscous discs can lead to expanding binary orbits.

For our simulations we assumed a coplanar configuration where the disc and binary lie the same plane.
This is motivated by the observed coplanar distribution of detected circumbinary planets and the fact that
the inner disc will align towards the binary plane over time, see e.g.\ \cite{2018Pierens}.
Additionally, we restricted ourselves to a 2D treatment and neglected the vertical disc structure. Only this simplification
allowed us to perform the simulations over viscous timescales.
The aligned 3D SPH-model in \cite{2020Hirsh} produce comparable cavity structures and sizes as 2D models, and
therefore we expect that the restriction to a 2D-setup can produce realistic results.

In our models we kept the binary stars on a fixed orbit. \cite{2019Munoz} carefully evaluated the binary migration rate $\dot{a}/a$ in a dynamic model and find to be on the order of 2 time the mass accretion rate for the $\alpha=h=0.1$ case. For even smaller viscosities it will migration will be considerably slower, accordingly. We tested $\dot{a}/a$ in the disc structure caused by the gravitational term $\dot{J}_\mathrm{grav}$ and found no deviations with or without "live" binaries in the simulations for various viscosities. 
We did not include the momentum transfer onto the binary stars in this study, as it is not computationally feasible to create models with very lower viscosity including the motion of the binaries directly, due to the very long convergence time needed.

In Eq.~(\ref{eqn:adot_bin3}) we omitted the eccentricity term as we only looked at circular binaries. 
Considering also the change in eccentricity would lead to a second variable to solve for and the resulting equation would not be solvable with just angular momentum conservation.
The additional property to look for is the momentum change of the binary induced by the mass accretion
which can only be computed if the stars are contained within the computational domain, which is not feasible for the current models.

As circular binaries represent that configuration with maximum angular momentum for a given semi-major axis we might expect as a consequence that
positive angular momentum transfer will always keep the orbit circular. Models with live binaries show indeed that the induced eccentricity remains indeed
very small for initially circular orbits \citep[see for example][]{2019Munoz,2020A&A...641A..64H}.

By using the simple prescription of a locally isothermal environment, the model is simple but scalable. This allows for general parameter studies for diverse astronomical discs.
In case a specific system is examined, a radiative model that includes viscous heating and possibly irradiation can be more physical. In \cite{2019Kley} we tested viscously heated and radiatively cooled circumbinary discs using the parameters of the Kepler-38 and Kepler-16 systems and found that the locally isothermal assumption with a flat aspect ratio between $4\%$ and $5\%$ for the inner disc provides a good approximation to the more realistic case.

In the models with low viscosities the accretion of mass and angular momentum becomes very small due to the large cavity size and long viscous evolution time scales.
This also implies that the occurring temporal variations in the accretion rate result in larger fluctuations of the measured $j_\mathrm{s}$.
Therefore, to obtain the mean $j_\mathrm{s}$ we have to calculate a rolling average over larger time interval, up to $100\,T_\mathrm{bin}$, or 1000 outputs if obtained from
individual snapshot. Still $j_\mathrm{s}$ for $\alpha=10^{-3}$ varies over $\pm 0.2$. Nevertheless, our result that for low viscosity thin discs the binary orbits shrink
seems to be robust as the decline of $j_\mathrm{s}$ is much larger than these fluctuations. Since for small values of viscosity and $H/r$ numerical effects may affect the results
we ran additional simulations using higher grid resolutions and in addition an infinite disc setup. The results, presented in \ref{subsec:infinite}, show that $\js$ increases 
for higher resolution but not enough to change our conclusion.

We do consider the binary as point like gravitational sources. However, the planet hosting Kepler binary systems are close binaries that in addition to the disc's impact are influenced by tidal interactions between the stars. In a theoretical study of Kepler 47b \citep{2020Graham}, the binary stars have roughly a semi-major axis change of $0.5\%$ over $10^9\,$years. This would be a much slower change than the effect we see from the disc. Therefore we can safely assume that the disc is dominant in changing the semi-major axis even for such kinds of close binaries.
\\\\
Overall, the regime of binary expansion extends to a much wider parameter space than previously anticipated which will have consequences with respect to the
orbital evolution of binary black holes prior to merger or the evolution of stellar binaries in their formation phase.
These results for very low disc thicknesses and viscosities are somewhat preliminary as our numerical resolution may not have been high enough to cover these cases properly.
Additional simulations using higher numerical resolution suggest that $\js$ increases, which could imply that the range of binary expansion will be somewhat larger, not changing our conclusion of shrinking binaries for low $\alpha$ and small $H/r$.

Future simulations need to address more realistic physical conditions where radiative effects are included.
The geometry needs to be extended to full 3D to verify the thin disc approach as recent simulations of massive embedded planet in discs point towards
a discrepancy between 2D and 3D simulations with respect to the eccentricity of the outer disc \citep{2021ApJ...906...52L}.

\begin{acknowledgements}
During the course of the revision of this paper, our colleague Willy Kley passed away unexpectedly. We want to express our sincere gratitude to him, who was not only our mentor, but also a dear and supportive friend. Thank you, Willy!
Anna Penzlin and Hugo Audiffren were both funded by grant 285676328 
from the German Research Foundation (DFG).
The authors acknowledge support by the High Performance and
Cloud Computing Group at the Zentrum für Datenverarbeitung of the University
of Tübingen, the state of Baden-Württemberg through bwHPC and the German
Research Foundation (DFG) through grant no INST 37/935-1 FUGG.
Some of the plots in
this paper were prepared using the Python library matplotlib \citep{Hunter:2007}.
\end{acknowledgements}

\bibliography{cb}
\bibliographystyle{aa}
\newpage
\begin{appendix}

\section{Code comparison}
\label{sec:A_code}
In order to validate complex hydrodynamical simulations it is very helpful, if not mandatory, to perform simulations using completely different numerical approaches on the same physical problem.
Even though in this paper we have compared the two grid codes for a wider parameter range it is usually prohibited to do this for all different choices of
parameter. Hence, we have chosen our standard model and performed simulations for three different codes: The finite volume codes \texttt{PLUTO} and \texttt{RH2D},
and a SPH code. 
 2007Mignone,  
The first code used is a full GPU version of the hydrodynamical part of the \texttt{PLUTO}-code \citep{2007Mignone} as described in \citet{2017Thun}.
It is a Riemann-solver based finite volume code, where we used the van Leer limiter for the reconstruction and a second order time-stepping.
As an alternative grid code we used \texttt{RH2D} which is a classic second-order upwind code working on a staggered grid \citep{1989A&A...208...98K,1999MNRAS.303..696K},
using again the van Leer limiter for the slopes. 

\begin{figure}[t]
    \centering
    \includegraphics[width=0.45\textwidth]{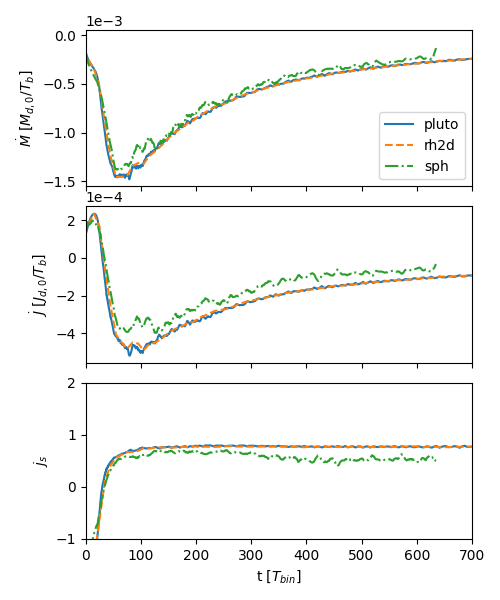}
    \caption{The mass accretion, angular momentum change of the disc and resulting $j_\mathrm{s}$ over time for our two grid codes \texttt{PLUTO} (blue, solid),
   \texttt{RH2D} (orange, dashed), and the SPH code \texttt{miluphcuda} (green, dash-dotted).
   }
    \label{fig:comp_m_l}
\end{figure}
For the SPH simulations we 
used the GPU code \texttt{miluphcuda} \citep{2016A&A...590A..19S, 2020A&C....3300410S}. In the standard model, we chose the same inner and outer boundary radius of the computational domain as the grid simulations, $1a$ to $40a$. Particles that cross the inner or outer boundary during the simulation are taken out of the computation. All particles have the same mass and are distributed in the beginning to represent the initial density profile. The initial smoothing lengths of the particles are chosen to ensure an overall constant number of interactions and to match the density profile. We apply variable smoothing and integrate the smoothing length using the standard approach (in 2D)

\begin{align}
\frac{\mathrm{d}h_\mathrm{SPH}}{\mathrm{d}t} = - \frac{h_\mathrm{SPH}}{2} \nabla \cdot \vec{v},
\end{align}
where $h_\mathrm{SPH}$ is the smoothing length and $\vec{v}$ the velocity. The code solves the Navier-Stokes equations following the scheme by \cite{1994ApJ...431..754F}, where the kinematic viscosity is given by the $\alpha$ model (see above). Additionally, an artificial bulk viscosity is applied following \cite{2004A&A...418..325S} and the XSPH algorithm for additional smoothing of the particle velocities is used \citep{MONAGHAN19891}. The standard setup includes initially \num{500000} particles.

\begin{figure}[t]
    \centering
    \includegraphics[width=0.45\textwidth]{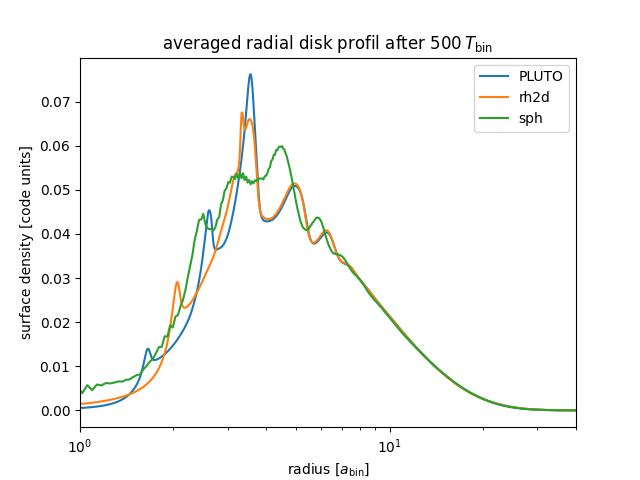}
    \caption{The azimuthally averaged surface density profile at 500 binary orbits of our two grid codes \texttt{PLUTO} (blue) and \texttt{RH2D} (orange), and the SPH code \texttt{miluphcuda} (green).
   }
    \label{fig:comp_sigma}
\end{figure}

All three simulations start from the same initial setup and are run for at least $600\,T_\mathrm{bin}$.
In Fig.~\ref{fig:comp_m_l} we show the time evolution for mass accretion and angular momentum change of the disc, and the specific angular momentum transfer $j_\mathrm{s}$.
The results for the \texttt{RH2D} code are identical to those displayed in Fig.\,\ref{fig:fluxesc_standard} above.
The two grid based codes show an excellent agreement for all three variables throughout the whole evolution.
The SPH code has a slightly lower mass and angular momentum transfer resulting in a lower $j_\mathrm{s}$
but it lies still well above $\jscrit$. This difference is possibly caused by the different treatment of the viscosity in SPH. In the very low density region
in the inner parts of the disc near the central binary there are fewer particles leading possibly to an inaccurate estimate of the viscosity.
In Fig. \ref{fig:comp_sigma} we compare the azimuthally averaged radial density profile after $500\,T_\mathrm{bin}$.
The results of all codes are almost perfectly aligned, except for highly dynamic spiral features in the density profile in the central regions.
At the inner boundary the SPH code has slightly higher density, 
but the position of the density maximum, the onset of the spiral features, and the density of in the outer disc match well between all models. 
In particular, it is worth noting that the viscous outward spreading of the initial torus is identical for all 3 codes, indicating
identical viscosity. We conclude that (at least for the standard model) the physical behaviour and conclusion about the binary evolution is robust.

\section{Estimating the accretion ratio}
\label{sec:appC_facc}

Here, we present briefly our initial studies in estimating the accretion factor $f$. For this purpose we use two different approaches.
First the grid-code \texttt{RH2D}, now in a Cartesian setup, and secondly the particle based SPH-code \texttt{miluphcuda}.
  
\subsection{Cartesian grid simulation}
\label{subsec:f}

To calculate the orbital evolution of the binary an estimate of the accretion factor, $f$, i.e. the mass accretion ratio onto the two stars is required.
As this is not possible for a cylindrical grid which has an inner hole at $r_\mathrm{min}$, we ran in addition Cartesian simulations using the two grid codes.
Here, we cover the a domain of $[-20:20,-20:20]$ for the standard model with a Cartesian grid using different number of grid cells, ranging from $1200 \times 1200$
up to $4800 \times 4800$.
In deviation from the standard model we use here a constant (dimensionless) viscosity of $\nu = 1.41 \cdot 10^{-3}$, to avoid problems
in defining suitable disc thicknesses in the vicinity of the two stars. 
The initial density distribution is identical to the standard model, using $r_\mathrm{in} = 2.5$ and $r_\mathrm{out}=6$.
For the stellar potentials we use a Plummer-type smoothing with $r_\mathrm{sm1} = 0.15\,M_1$ and $r_\mathrm{sm2} = 0.15\, M_2$ for the two stars.
For the sink radii we use 2/3 $r_\mathrm{sm}$ for the two stars.
Inside the sink radii of the two stars mass is taken out at with a half-emptying time of $\tau_{1/2} \sim 7\,T_\mathrm{bin}$.

\begin{figure}[t]
    \centering
    \includegraphics[width=0.45\textwidth]{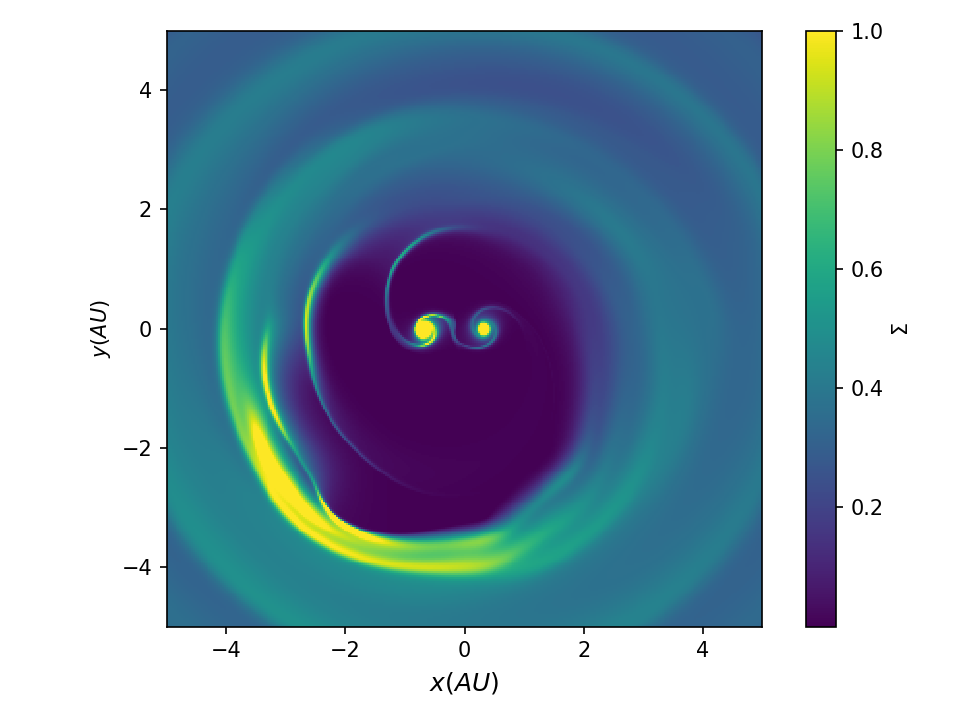}
    \caption{Density structure in the central region for the \texttt{RH2D} model with $3400\times3400$ cells with the parameters of the standard model after 500 binary periods.
   }
    \label{fig:density_rh2d}
\end{figure}

\begin{figure}[t]
    \centering
    \includegraphics[width=0.45\textwidth]{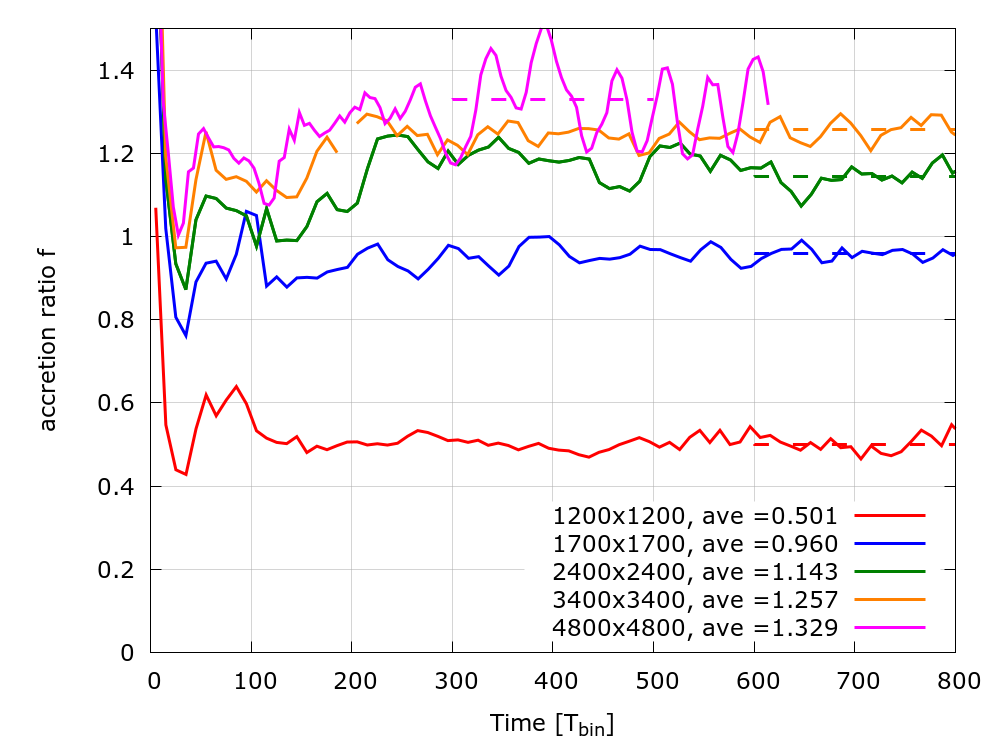}
    \caption{The stellar mass accretion ratio, $f = \dot{M}_2/\dot{M}_1$, onto the two stars, for Cartesian simulations with different grid resolutions.
    Shown is the time evolution with the corresponding averages, taken over the time interval indicated by the dashed lines.
   }
    \label{fig:facc-time}
\end{figure}

The 2D density distribution of such a Cartesian simulation is displayed in Fig.~\ref{fig:density_rh2d} for the standard model
after a simulated time of $500\,T_\mathrm{bin}$.
Despite the accretion of mass the circumstellar discs are still visible around both stars, implying that the
relatively short $\tau_{1/2}$ is not effective enough to get rid of the circumstellar discs entirely.
From such simulations we can obtain the mass accretion rates on the two stars
and the important accretion ratio, $f$. We did not monitor in these simulations the angular momentum transfer to the stars, and leave this to future
studies.

The measured accretion ratio for $q=0.5$ onto the stars is displayed in Fig.~\ref{fig:facc-time} for different grid resolutions.
The models settle to an equilibrium at about 250 to $300\,T_\mathrm{bin}$ after which the accretion ratios remain approximately constant.
Increasing the grid resolution increases the accretion factor, and the results converge for the standard model against a value $f_\mathrm{lim} \approx 1.37$.
As a test for the accretion procedure, we performed the same simulation for a model with two equal mass stars ($q=1,  \alpha=0.1$, and $h=0.1$)
and found a mass accretion ratio $f=1$, meaning equal accretion onto both stars, as expected.

\subsection{SPH-Miluphcuda}

For the SPH simulations conducted in this project we define individual accretion radii around the stars, which are scaled with the corresponding mass. Whenever a particle crosses the accretion radius of a star, we monitor its mass, velocity and accretion time and remove it from the simulation. However, we chose the accretion radii small enough to prevent overlapping, i.e.\ $r_\mathrm{acc1} = 0.1\,M_1$ and $r_\mathrm{acc2} = 0.1\, M_2$.

\begin{figure}[t]
    \centering
    \includegraphics[width=0.45\textwidth]{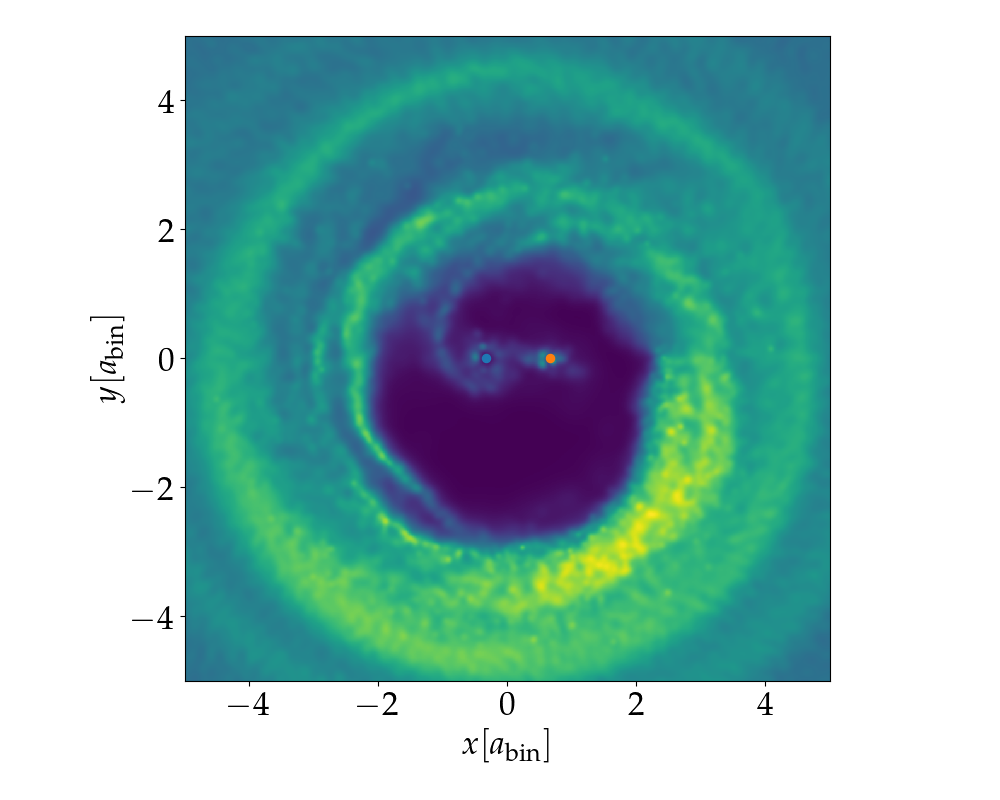}
    \caption{Density structure in the central region for the SPH model with the parameters of the standard model using \num{5e5} particles after 500 binary periods. For the colour scaling see Fig.~\ref{fig:density_rh2d}.
   }
    \label{fig:density_sph}
\end{figure}

\begin{figure}[t]
    \centering
    \includegraphics[width=0.45\textwidth]{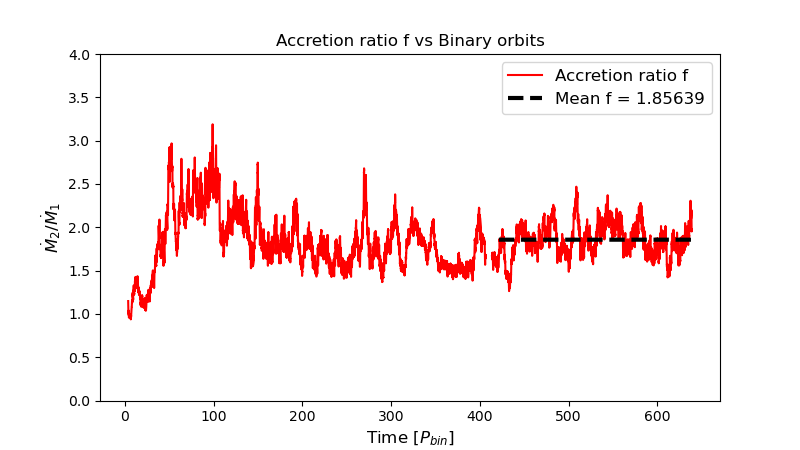}
    \caption{Accretion ratio evolution for the SPH model shown in Fig.~\ref{fig:density_sph}.
   }
    \label{fig:frat_sph}
\end{figure}

The 2D density distribution of the model after a time $t=500\, T_\mathrm{bin}$ is shown in Fig.~\ref{fig:density_sph}. The overall structure is
very comparable to the results of the cylindrical and Cartesian grid models with similar cavity size. The density distribution was calculated
by interpolating the particles to a cylindrical grid such that the same analysis tools could be used as for the grid codes.
The SPH accretion ratio is plotted in Fig.~\ref{fig:frat_sph}, and the mean value $f_\mathrm{SPH,mean} \approx 1.86$ is computed over the last 200~orbits.

In our parameter study we used an average $f = 1.6$ for the relative mass accretion rates in calculating the critical
$\jscrit$, assuming that a similar value will be taken for other values of $\alpha$ and $h$. 
This value is in good agreement with the recent study by \citet{2021ApJ...921...71D}
who performed a more detailed analyses of mass and angular momentum accretion using different numerical
recipes, and with \cite{2020ApJ...889..114M} who used the moving mesh code AREPO, and found a slightly higher accretion ratio
of $\sim 1.7$ after about 400 orbits for similar disc and binary parameter (see their Fig.~9).

\end{appendix}
\end{document}